\documentclass[11pt,reprint,english,aps,prc,showpacs,letter,longbibliography]{revtex4-1}

\usepackage[ bookmarks,
             bookmarksopen = true,
             bookmarksnumbered = true,
             linktocpage,
             colorlinks = true,
             linkcolor = blue,
             urlcolor  = blue,
             citecolor = blue,
             anchorcolor = green,
             hyperindex = true,
             hyperfigures]{hyperref}

\usepackage{dcolumn}
\usepackage{bm}
\usepackage{float}
\usepackage{amsmath}
\usepackage{cleveref}

\usepackage{booktabs}    
\usepackage{multirow}    

\usepackage{color}
\usepackage{graphicx}
\definecolor{dgreen}{cmyk}{1.,0.,1.,0.2}        
\definecolor{orange}{cmyk}{0.,0.353,1.,0.}    

\newcommand{\trento}{T\raisebox{-0.5ex}{R}ENTo}
\newcommand{\trentoDDD}{T\raisebox{-0.5ex}{R}ENTo\nobreakdash-3D}

\begin{document}

\title{Study the Longitudinal Entropy Deposition using d+Au Collision}%

\author{Zhu Meng}
\affiliation{Key Laboratory of Quark and Lepton Physics (MOE) and Institute of Particle Physics, Central China Normal University, Wuhan 430079, China}
\author{Weiyao Ke}
\affiliation{Key Laboratory of Quark and Lepton Physics (MOE) and Institute of Particle Physics, Central China Normal University, Wuhan 430079, China}
\author{Long-Gang Pang}
\email{lgpang@ccnu.edu.cn}
\affiliation{Key Laboratory of Quark and Lepton Physics (MOE) and Institute of Particle Physics, Central China Normal University, Wuhan 430079, China}


\date{\today}%

\begin{abstract}

Relativistic hydrodynamics successfully describes bulk observables in symmetric heavy-ion collisions, but struggles to reproduce charged-particle rapidity distributions in asymmetric systems such as d+Au collisions. To address this challenge, we introduce two key improvements to the initial-state modeling: sampling deuteron configurations from an ab initio wavefunction, and developing a new longitudinal entropy deposition model that incorporates a transverse entropy deposition coefficient $\beta$ and a rapidity loss term scaling with the number of binary collisions $n_{\rm BC}$. Using the (3+1)-dimensional viscous hydrodynamic model CLVisc coupled with the SMASH afterburner, we simulate d+Au collisions at $\sqrt{s_{\rm NN}} = 200$ GeV and successfully reproduce the experimental charged-particle pseudorapidity distributions across five centrality classes with $\beta = 0.35$, as well as the transverse momentum spectra and anisotropic flow $v_n$. The entropy deposition coefficient $\beta$ and the $n_{\rm BC}$-dependent rapidity loss are found to play crucial roles in achieving this agreement. Furthermore, this longitudinal entropy deposition framework demonstrates excellent universality, as validated in p+Au, $^3$He+Au, and Au+Au collisions. Our entropy deposition mechanism could be widely applied to recent light-nucleus collisions such as O+O, Ne+Ne, and asymmetric systems like Pb+Ne at LHC energies, thereby better constraining the nuclear structure of light nuclei through an improved longitudinal description.

\end{abstract}

\maketitle

\section{Introduction}

Relativistic heavy-ion collisions deconfine quarks and gluons from nucleons and create a new state of nuclear matter, called quark gluon plasma (QGP), which is the same as the primordial nuclear matter created in the early universe, about 1 microseconds after the big bang ~\cite{Lee:1974ma,Shuryak:1980tp,Busza:2018rrf,STAR:2005gfr}.
More controversially, the observation of long-range near-side ridge structures in di-hadron correlation functions in high-multiplicity in d+Au collisions at RHIC~\cite{PHENIX:2013ktj,STAR:2015kak} has been suggested as a potential signature of QGP formation in small systems
However, relativistic hydrodynamics, which successfully describes the collective expansion in large systems, still struggles to reproduce the charged-particle rapidity distributions in asymmetric d+Au collisions.




The charged particle pseudorapidity distribution $dN_{\rm ch}/d\eta$ in asymmetric d+Au collisions serves as an important observable for testing longitudinal entropy deposition mechanisms in different initial condition theoretical models.
Experimentally, this distribution has been extensively measured by various experiments~\cite{PHOBOS:2004fzb,BRAHMS:2004taz,STAR:2004ggj,STAR:2007poe}.
Phenomenologically, several approaches have been proposed to describe it, including the AMPT model~\cite{Lin:2001yd,Lin:2003ah}, the wounded-quark model~\cite{Bialas:1976ed,Barej:2017kcw,Rohrmoser:2018shp,Barej:2019xef}, the HIJING model~\cite{Deng:2010mv}, and Tsallis thermodynamics~\cite{Tao:2023kcu}. While these models provide reasonable descriptions of the multiplicity distributions, they typically rely on phenomenological parameters with limited connection to the underlying QGP dynamics.
On the theoretical side, initial condition models coupled with relativistic hydrodynamics have achieved great success in symmetric systems such as Au+Au and Pb+Pb~\cite{Pang:2018zzo}, where a common approach is to extend the two-dimensional transverse entropy density to three dimensions by multiplying a one-dimensional longitudinal deposition function. However, this simple extension, while effective for symmetric collisions, performs poorly for asymmetric d+Au collisions.
Several studies~\cite{Shen:2016zpp,Shen:2020jwv,Zhao:2022ugy} have attempted to combine three-dimensional initial condition generation with (3+1)-dimensional hydrodynamic models such as MUSIC to simulate the $dN_{\rm ch}/d\eta$ distribution in d+Au collisions, but a complete description across the full pseudorapidity and centrality ranges remains lacking. The \trentoDDD~\cite{Moreland:2014oya,Moreland:2018gsh,Soeder:2023vdn} initial condition model generates three-dimensional initial conditions via Bayesian analysis and performs well for Au+Au collisions~\cite{Zhu:2026vql,Jiang:2025sad}, yet it still struggles to fully describe d+Au collisions.

Nuclear structure plays a crucial role in shaping the initial conditions of heavy-ion collisions, with imprints that persist into final-state observables. 
In particular, nuclear deformation~\cite{Ryssens:2023fkv,Lu:2023fqd,Giacalone:2021udy,Jia:2021tzt,Schenke:2014tga,STAR:2024wgy}, $\alpha$-clustering in light nuclei~\cite{Shi:2021far,Freer:2017gip,Wang:2021ghq,Summerfield:2021oex,Ding:2023ibq,Rybczynski:2017nrx}, isobaric shape differences~\cite{Bhatta:2023cqf,Jia:2022qgl,Zhang:2021kxj,Xu:2021uar}, and the neutron skin thickness~\cite{Abrahamyan:2012gp,Giacalone:2023cet,Piekarewicz:2010fa} have all been shown to affect collective flow observables. 
At the microscopic level, nucleon-nucleon correlations and the finite size of nucleons introduce additional fluctuations that are sensitive to final-state observables~\cite{Alvioli:2009ab,Alvioli:2011sk,Broniowski:2010jd,Xu:2025cgx,Wang:2026pcx,Kozlov:2014fqa}. 
For the specific case of the deuteron, its intrinsic structure as a loosely bound proton-neutron system introduces unique two-body correlations that are not captured by simple independent-nucleon sampling and may significantly influence the initial geometry in d+Au collisions.

The present work focuses on two key aspects of d+Au collisions. 
First, we investigate the effect of the deuteron nuclear structure by employing first-principles wavefunctions that incorporate both nucleon-nucleon correlations and the interference between S- and D-wave components, rather than treating the deuteron as an isotropic system.
Second, we examine the mechanism of initial longitudinal entropy deposition through a newly constructed three-dimensional entropy density model that includes an entropy deposition coefficient $\beta$ and a binary-collision-dependent rapidity loss $\Delta\eta_s(n_{\rm BC})$, capturing baryon stopping phenomena in high-energy collisions~\cite{Videbaek:1995mf,Blume:2007kw,Zhou:2009yt,Busza:1983rj,Garcia-Montero:2024jev}. 
This model goes beyond the simple factorization approach commonly used in symmetric systems, where transverse and longitudinal components are treated independently.

This initial condition is implemented in the (3+1)-dimensional viscous hydrodynamic model CLVisc~\cite{Pang:2018zzo} to simulate the QGP evolution, followed by hadronization via the SMASH afterburner~\cite{SMASH:2016zqf}. By calculating the final-state charged particle multiplicity distribution $dN_{\rm ch}/d\eta$, transverse momentum spectra, and anisotropic flow $v_n$, and comparing them with experimental data, we validate our framework. To further test its universality, we apply the same framework to p+Au $^3$He+Au and Au+Au collisions, demonstrating that the model also describes symmetric systems well. Our results provide a unified description of both symmetric and asymmetric collision systems, offering deeper insights into the interplay between projectile nuclear structure and longitudinal entropy deposition mechanisms.

The paper is organized as follows. 
Section~\ref{sec:method} presents the methodological framework, including the sampling of nucleon distributions from deuteron wavefunctions, the construction of the three-dimensional initial condition with entropy deposition coefficient $\beta$ and binary-collision-dependent pseudorapidity loss, and the hydrodynamic simulations with CLVisc and the SMASH afterburner. Section~\ref{sec:results} presents the results, covering the effects of deuteron nuclear structure, the $dN_{\rm ch}/d\eta$ distributions, comprehensive observables from CLVisc+SMASH, and the universality test. Section~\ref{sec:summary} provides a summary and discussion.

\section{Method}\label{sec:method}

\subsection{Sample Nucleons from Deuteron }

In general, heavy atomic nuclei consist of a large number of nucleons, whose spatial distribution is commonly described by the Woods-Saxon distribution. The nucleon density $\rho_A$ is given by
\begin{equation}\label{equation:W-S_distribution}
    \rho_A(r) = \frac{\rho_0}{1 + e^{(r - R_A)/d_A}},
\end{equation}
where $\rho_0 \approx 0.17 \,\mathrm{fm}^{-3}$ is the central nucleon density, $R_A$ is the nuclear radius, and $d_A$ is the surface thickness parameter.


In contrast, the deuteron nucleus in the \trento{} model uses the Hulth\'en wavefunction (HWF), which provide an analytical description of the deuteron ground state~\cite{DONNACHIE1964128,Alver:2008aq,Hulthen}:
\begin{equation}\label{equation:psi_HWF}
    \psi_{\rm HWF}(r) = \sqrt{\frac{\alpha \beta (\alpha + \beta)}{2\pi(\alpha - \beta)^2}} \, \frac{e^{-\alpha r} - e^{-\beta r}}{r},
\end{equation}
where $r$ is the proton-neutron separation distance, $\alpha$ and $\beta$ are parameters chosen to reproduce the deuteron binding energy and charge radius, typically $\alpha = 0.228 \, \mathrm{fm}^{-1}$ and $\beta = 1.18 \, \mathrm{fm}^{-1}$. The corresponding probability density is then given by $\rho_{\rm HWF}(r) = |\psi_{\rm HWF}(r)|^2$. Since $\rho_{\rm HWF}(r)$ depends only on the radial coordinate $r$, the joint probability density can be factorized into radial and angular components:
\begin{equation}\label{equation:separate_P}
    \begin{aligned}
        p(r) &= 4\pi r^2 \rho_{\rm HWF}(r), \quad r \in [0, \infty), \\
        p(\theta) &= \frac{\sin \theta}{2}, \quad \theta \in [0, \pi], \\
        p(\phi) &= \frac{1}{2\pi}, \quad \phi \in [0, 2\pi).
    \end{aligned}
\end{equation}
Here $p(r)$ is the radial probability density satisfying $\int_0^\infty p(r) dr = 1$, while $p(\theta)$ and $p(\phi)$ are the angular probability densities for the polar and azimuthal angles, respectively.

According to Eq.~\eqref{equation:psi_HWF}, the probability density depends only on the radial coordinate $r$, with $r$, $\theta$, and $\phi$ being mutually independent variables. This enables separate sa
mpling from the distributions $p(r)$, $p(\theta)$, and $p(\phi)$ defined in Eq.~\eqref{equation:separate_P}, from which the spatial coordinates are constructed. The sampling results show that $\theta$ follows a $\sin\theta$ distribution, while $\phi$ is uniformly distributed.

A more realistic scenario involves the deuteron wave function (DWF), which can be expressed as a superposition of the $^3S_1$ and $^3D_1$ states \cite{BlattWeisskopf1958, Zhaba:2017syr}:
\begin{equation}\label{equation:psi_d}
    \Psi_{d} = \psi_{S} + \psi_{D} = \frac{u(r)}{r} \mathcal{Y}_{101}^1(\theta,\phi) + \frac{w(r)}{r} \mathcal{Y}_{121}^1(\theta,\phi),
\end{equation}
where $u(r)$ and $w(r)$ are the radial wave functions for the S- and D-states, respectively. The functions $\mathcal{Y}_{JLS}^M(\theta,\phi)$ are vector spherical harmonics characterized by orbital angular momentum $L$, spin $S$, total angular momentum $J = L + S$, and its projection $M$. For the deuteron, $S = 1$ and $J = M = 1$. Thus, $\mathcal{Y}_{101}^1$ and $\mathcal{Y}_{121}^1$ correspond to the $L = 0$ (S-state) and $L = 2$ (D-state) contributions, respectively.

The radial wave functions $u(r)$ and $w(r)$ are given by \cite{Zhaba:2015yxq, Zhaba:2017syr, Wiringa:1994wb}:
\begin{equation}\label{equation:radial_dwf}
    \left\{
        \begin{array}{l}
            u(r) = r^{3/2} \sum_{i=1}^{N} A_{i} \exp \left(-a_{i} r^{3}\right) \\\\
            w(r) = r \sum_{i=1}^{N} B_{i} \exp \left(-b_{i} r^{3}\right),
        \end{array}
    \right.
\end{equation}
where the coefficients $A_i$, $a_i$, $B_i$, and $b_i$ parametrize the radial deuteron wave function, adopted from the Argonne $\text{v}_{18}$ potential as tabulated in Table 11 of Ref.~\cite{Zhaba:2017syr}.

The vector spherical harmonics $\mathcal{Y}_{101}^1$ and $\mathcal{Y}_{121}^1$ can be expressed in terms of spherical harmonics and spin states~\cite{BlattWeisskopf1958}:
\begin{equation}\label{equation:vector_harmonics}
    \left\{
        \begin{array}{l}
            \mathcal{Y}_{101}^{1} = Y_{0,0} \, \chi_{1,1} \\\\
            \mathcal{Y}_{121}^{1} = \sqrt{\dfrac{6}{10}} Y_{2,2} \chi_{1,-1} - \sqrt{\dfrac{3}{10}} Y_{2,1} \chi_{1,0} + \sqrt{\dfrac{1}{10}} Y_{2,0} \chi_{1,1},
        \end{array}
    \right.
\end{equation}
where $Y_{L,m}$ are the spherical harmonics:
\begin{equation}\label{equation:spherical_harmonics}
    \left\{
    \begin{aligned}
        Y_{0,0} &= \sqrt{\frac{1}{4\pi}} \\
        Y_{2,2} &= \sqrt{\frac{15}{32\pi}} \sin^2\theta \, e^{2i\phi} \\
        Y_{2,1} &= -\sqrt{\frac{15}{8\pi}} \sin\theta \cos\theta \, e^{i\phi} \\
        Y_{2,0} &= \sqrt{\frac{5}{16\pi}} (3\cos^2\theta - 1),
    \end{aligned}
    \right.
\end{equation}
and $\chi_{1,m_s}$ are the spin-1 eigenstates satisfying $\chi_{1,m_s} \chi_{1,m'_s} = \delta_{m_s m'_s}$, explicitly given by:
\begin{equation}\label{equation:spin_states}
    \left\{
    \begin{aligned}
        &\chi_{1,1} = |\uparrow\uparrow\rangle, \\
        &\chi_{1,0} = \frac{1}{\sqrt{2}} (|\uparrow\downarrow\rangle + |\downarrow\uparrow\rangle), \\
        &\chi_{1,-1} = |\downarrow\downarrow\rangle.
    \end{aligned}
    \right.
\end{equation}

According to Eqs.~\eqref{equation:psi_d}–\eqref{equation:spin_states} and the fundamental principles of quantum mechanics, the wave function represents a probability amplitude, from which the spatial probability density of nucleons can be determined. Thus, the probability density is given by
\begin{equation}\label{equation:rho_D}
    \begin{aligned}
        \rho_{\rm DWF}(r,\theta)=&|\Psi_d(r,\theta,\phi)|^{2}\\
        =&\big|\frac{u(r)}{r}\mathcal{Y}_{101}^{1}+\frac{w(r)}{r}\mathcal{Y}_{121}^{1}\big |^2\\
        =& \left| \frac{w(r)}{r} \sqrt{\frac{6}{10}} Y_{2,2} \right|^2 + \left| -\frac{w(r)}{r} \sqrt{\frac{3}{10}} Y_{2,1} \right|^2 \\
        & + \left| \frac{u(r)}{r} Y_{0,0} + \frac{w(r)}{r} \sqrt{\frac{1}{10}} Y_{2,0} \right|^2\\
        =& \frac{1}{r^2} \left[ u^2(r)  |Y_{0,0}|^2 +  2 u(r) w(r) \sqrt{\frac{1}{10}}  Y_{0,0} Y_{2,0}  \right. \\
        & +  \left. w^2(r) \left( \frac{6}{10} |Y_{2,2}|^2 + \frac{3}{10} |Y_{2,1}|^2 + \frac{1}{10} |Y_{2,0}|^2 \right) \right] \\
        =& \frac{1}{4 \pi r^{2}} \left[ u^{2}(r) + \frac{u(r) w(r)}{\sqrt{2}} \left( 3 \cos^{2} \theta - 1 \right) \right. \\
        &+ \left. \frac{w^{2}(r)}{4} \left( 5 - 3 \cos^{2} \theta \right) \right]. 
    \end{aligned}    
\end{equation}

From Eq.~\eqref{equation:rho_D}, a coupling between the variables $r$ and $\theta$ is observed. This coupling arises because both $\mathcal{Y}_{101}^{1}$ and $\mathcal{Y}_{121}^{1}$ contain $\chi_{1,1}$ components, leading to interference terms such as $Y_{0,0}Y_{2,0}$ in $\rho_{\rm DWF}$. This interference produces constructive and destructive effects along specific angular directions, resulting in a complex angular distribution of nucleons within the deuteron nucleus.

\begin{figure*}[htp]
    \centering
    \includegraphics[width=0.8\textwidth]{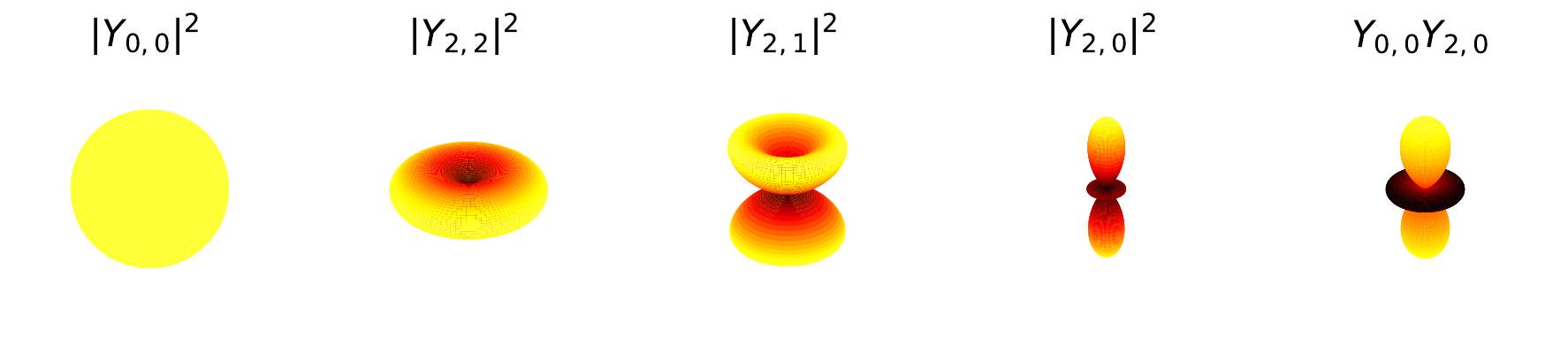}
    \caption {(Color online) Angular probability distributions of the squared magnitudes of spherical harmonics $Y_{l,m}$ in three-dimensional space for different quantum numbers $l,m$ and their product. $|Y_{0,0}|^2$: the simplest spherical harmonic distribution ($l=0$, $m=0$), uniformly distributed in all directions, forming a smooth spherical surface with no directional dependence. $|Y_{2,0}|^2$, $|Y_{2,1}|^2$, $|Y_{2,2}|^2$, and $Y_{0,0}Y_{2,0}$: highly symmetric and directional multipolar distributions, highlighting the angular complexity of this state. For the first four panels, color indicates the relative magnitude of $|Y_{l,m}|^2$, transitioning from dark red (low probability density) to yellow (high probability density). For the last panel ($Y_{0,0}Y_{2,0}$), black regions represent negative values. Blank regions correspond to zero probability density.}
    \label{fig:spherical_harmonic}
\end{figure*}

Fig.~\ref{fig:spherical_harmonic} illustrates the angular distributions of the spherical harmonics, comparing the isotropic S-wave ($|Y_{0,0}|^2$) with the anisotropic D-wave components ($|Y_{2,m}|^2$ with $m=2,1,0$) and their quantum interference ($Y_{0,0}Y_{2,0}$). While preserving azimuthal ($\phi$) rotational symmetry, the $Y_{0,0}Y_{2,0}$ interference term significantly modulates the polar angle ($\theta$) distribution: constructive interference dominates at the poles ($\theta = 0,\pi$),  enhancing the probability density. As $\theta$ approaches $\pi/2$, the constructive interference gradually weakens, eventually transitioning to destructive interference near $\theta=\pi/2$, which suppresses the probability density. The degree of enhancement and suppression is determined by the coefficient of the interference term.
This characteristic modulation pattern provides direct visual confirmation of the $\theta$-dependent features observed in our sampling analysis.

\begin{figure}[htp]
  \includegraphics[width=0.45 \textwidth]{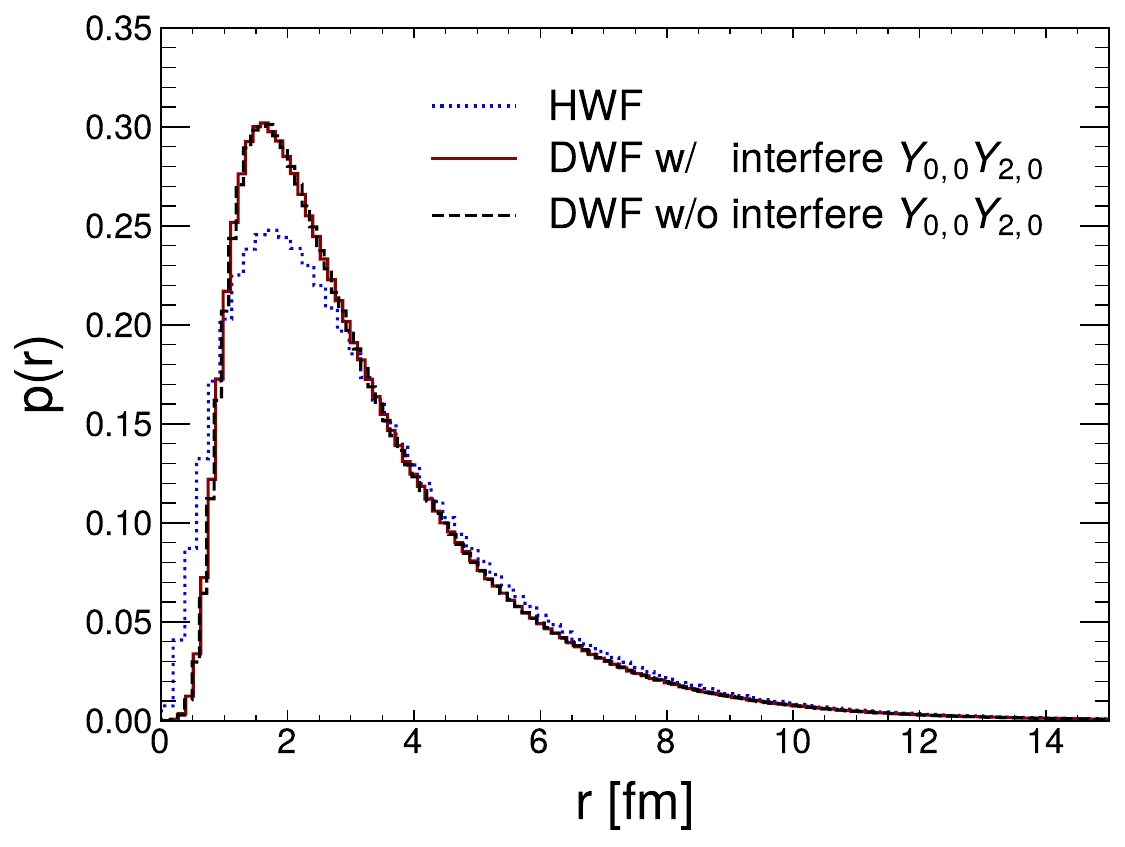}\\
  \includegraphics[width=0.45 \textwidth]{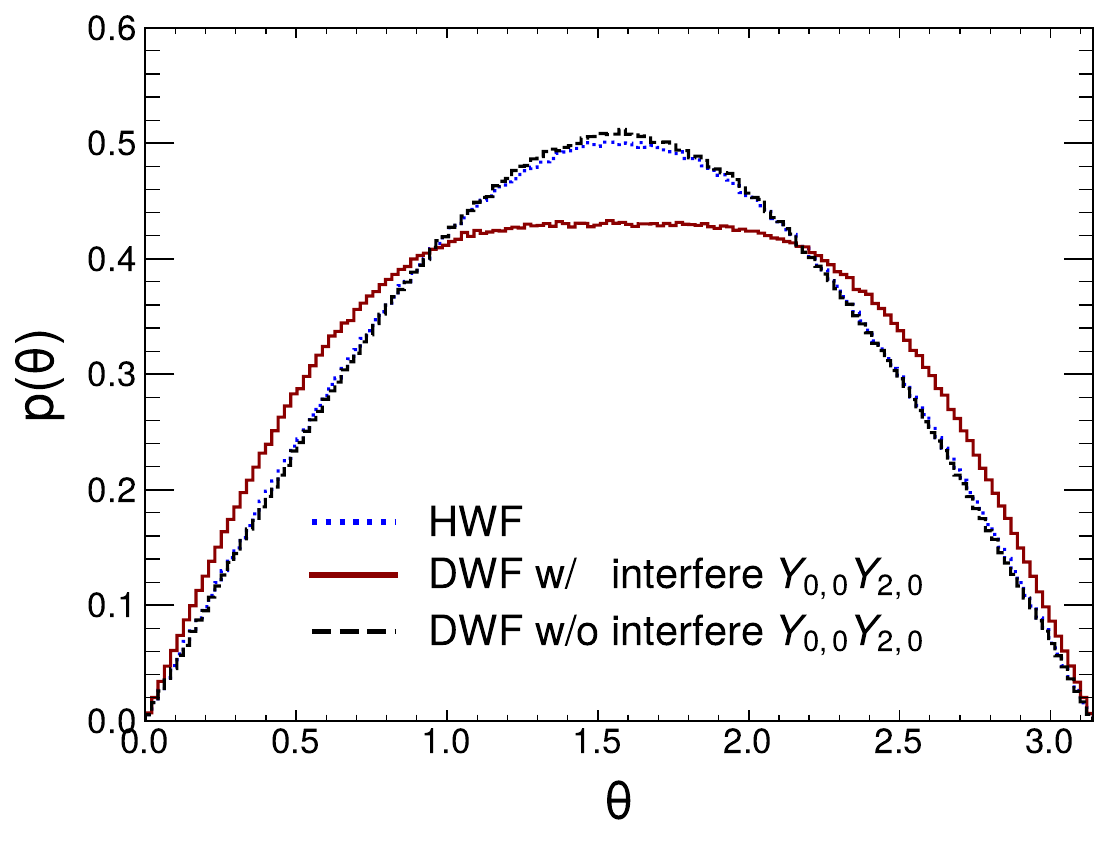} 
  \caption{(Color online)The upper and lower panels respectively depict the probability density distributions of $r$ and $\theta$ sampled from different nuclear structures: HWF, DWF, and DWF without the $Y_{00}Y_{20}$ term. The blue dashed curve represents the sampling results obtained using the HWF structure, while the red solid and black dashed curves correspond to the results sampled from the DWF structure with (w/) and without (w/o) the $Y_{00}Y_{20}$ term, respectively.}
  \label{fig:r_theta_phi_dis_in_DWF_and_HWF}
\end{figure}

Similar to the HWF, we define the joint probability density $p(r,\theta)$ and the marginal distribution $p(\phi)$ for the DWF as follows:
\begin{equation}\label{equation:P_DWF}
    \begin{aligned}
        &p(r,\theta) = 2\pi r^2 \sin\theta \, \rho_{\rm DWF}(r), \quad r \in [0, \infty), \; \theta \in [0, \pi],  \\
        &p(\phi) = \frac{1}{2\pi}, \quad \phi \in [0, 2\pi).
    \end{aligned}
\end{equation}
By sampling $r$ and $\theta$ simultaneously from Eq.~\eqref{equation:P_DWF}, we show that the DWF yields more realistic distributions for the relative distance $r$ and the polar angle $\theta$.

Fig.~\ref{fig:r_theta_phi_dis_in_DWF_and_HWF} illustrates key distinctions in nucleon distributions among the HWF (blue-dotted line), DWF (red-solid line), and DWF without the $Y_{0,0}Y_{2,0}$ interference term (gray-dashed line). The upper panel displays the radial probability density $p(r)$. Both DWF configurations, with and without the interference term, exhibit a higher probability than the HWF around the peak region at $r \approx 1.8$ fm. The interference term has a minimal impact on the radial distribution for the two DWFs. The probability of two nucleons being closer than 1 fm is higher for the HWF than for the DWF, indicating the effect of short-range repulsion in the nuclear force. For $r > 4$ fm, the probability is also higher for the HWF than for the DWF. Overall, the fluctuations in the relative distance $r$ are significantly smaller in the DWF than in the HWF.

The lower panel compares the polar angle distribution $p(\theta)$. The DWF with S-D wave interference term $Y_{0,0}Y_{2,0}$ exhibits a flatter distribution around $\theta = \pi/2$ and higher probability near $\theta = 0$ and $\pi$, as compared with HWF as well as the DWF without the S-D wave interference term. Specifically, $Y_{0,0}Y_{2,0}$ introduces destructive interference near $\theta = \pi/2$, reducing probability density, and constructive interference near $\theta = 0$ and $\pi$, enhancing it. The azimuthal angle $\phi$ remains uniformly distributed in all cases (not shown), consistent with rotational symmetry. The interference phenomenon fundamentally distinguishes the DWF from the HWF.

The tensor force between two nucleons, dominated by one-pion exchange, introduces an angular dependence through the operator~\cite{Xi:Phys741}
\begin{equation}
    S_{12} = 3(\vec{\sigma}_1 \cdot \hat{r})(\vec{\sigma}_2 \cdot \hat{r}) - \vec{\sigma}_1 \cdot \vec{\sigma}_2,
    \label{eq:tensor_operator}
\end{equation}
where $\vec{\sigma}$ represents the Pauli spin operator and $\hat{r}=\vec{r}/r$ is the normalized relative position vector between nucleons. Under the $M=1$ condition (both nucleon spins aligned along the $z$-axis), its expectation value reduces to
\begin{equation}
    \langle S_{12} \rangle = 3\cos^2\theta - 1,    
\end{equation}
where $\theta$ is the angle between $\hat{r}$ and the $z$-axis.

The tensor potential is $V_{\text{Tensor}} = V_T(r) S_{12}$, with $V_T(r) < 0$ (attractive) for one-pion exchange. Therefore:
\begin{itemize}
    \item When $\hat{r}$ is parallel to the $z$-axis ($\theta = 0$ or $\pi$), $\langle S_{12} \rangle = +2$, yielding strong attraction. This configuration is energetically favored, leading to an enhancement of the sampled probability $P(\theta)$ at $\theta = 0$ and $\pi$.
    \item When $\hat{r}$ is perpendicular to the $z$-axis ($\theta = \pi/2$), $\langle S_{12} \rangle = -1$, yielding repulsion. This configuration is suppressed, leading to a depletion of $P(\theta)$ near $\theta = \pi/2$.
\end{itemize}

This explains why the sampled angular distribution deviates from a uniform $\sin\theta$ distribution. The enhancement along the spin direction and suppression in the perpendicular plane directly reflect the tensor-force-induced $S$-$D$ wave mixing in the deuteron wave function.

\subsection{Longitudinal Entropy Deposition in 3D initial condition}\label{subsec:3d-initial-condition}

\begin{figure*}
    \includegraphics[width=1.0\textwidth]{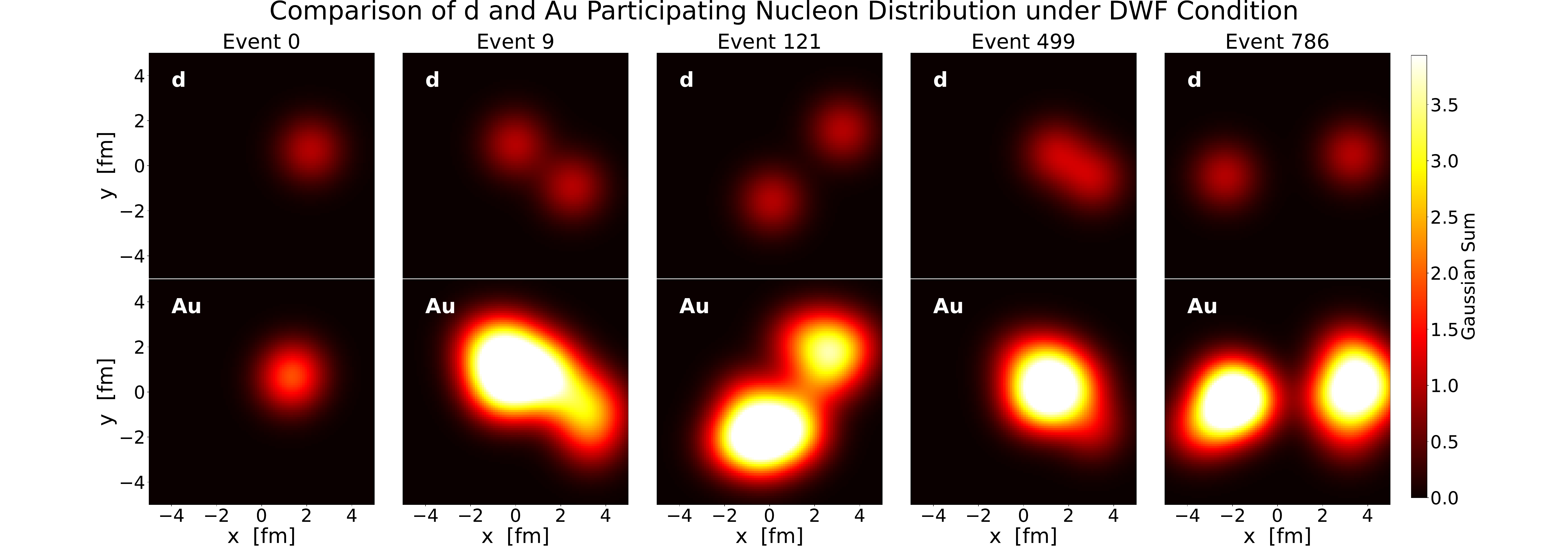}
    \caption{(Color online)Transverse plane distributions of participating nucleons from the deuteron (using the DWF structure) and gold nuclei (using the Woods-Saxon distribution) for five selected collision events (Event IDs: 0, 9, 121, 499, 786) in d+Au collisions. Color scale: summed contributions from all participating nucleons projected onto the $x$-$y$ plane. The transverse plane is discretized into a $200 \times 200$ grid from $-10$ fm to $10$ fm along both axes, with only the region from $-5$ fm to $5$ fm shown. Each nucleon's contribution is smeared with a Gaussian of width $\sigma_\perp = 0.5$ fm and is set to zero when $|\mathbf{x}_\perp - \mathbf{x}_i| > 5$ fm. First row: deuteron; second row: gold; each column corresponds to a distinct event. }
    \label{fig:hotmap_sub_2D_entropy}
\end{figure*}

The charged particle multiplicity distribution $\rm dN_{ch}/d\eta$ in asymmetric d+Au collisions serves as an excellent observable for probing longitudinal entropy deposition. In nucleus-nucleus collisions, the three-dimensional initial entropy density distribution is typically constructed by extending the MC-Glauber initial entropy density distribution along the longitudinal direction. This extension is commonly implemented using the following envelope function~\cite{Shen:2016zpp,Kozlov:2014fqa}:
\begin{equation}\label{equation:entropy_old}
    \begin{aligned}
        S(\mathbf{x}_{\perp},\eta_s) 
        &=F_L(\eta_s)\sum_{i=1}^{N_{\mathrm{part}}^{\mathrm{left}}}\rho(\mathbf{x}_\perp, \mathbf{x}_i)
        +F_R(\eta_s)\sum_{i=1}^{N_{\mathrm{part}}^{\mathrm{right}}}\rho(\mathbf{x}_\perp, \mathbf{x}_i),
    \end{aligned}    
\end{equation}
where $S(\mathbf{x}_{\perp},\eta)$ denotes the three-dimensional entropy density, $F_L(\eta_s)$ and $F_R(\eta_s)$ represent left-right symmetric longitudinal distribution functions. For d+Au collisions, $N_{\rm part}^{\rm left}$ and $N_{\rm part}^{\rm right}$ are numbers of participating nucleons in the left-going Au and right-going deuteron. The $\rho(\mathbf{x}_\perp, \mathbf{x}_i)$ characterizes the two-dimensional entropy density contribution from a single participating nucleon in the transverse plane. This distribution is defined as:
\begin{equation}\label{equation:single_entropy}
    \rho(\mathbf{x}_\perp, \mathbf{x}_i) = \exp\left(-\frac{(\mathbf{x}_\perp - \mathbf{x}_i)^2}{2\sigma_\perp^2}\right),
\end{equation}
where $\mathbf{x}_\perp$ indicates grid points in the transverse x-y plane, $\mathbf{x}_i$ denotes the transverse position of each participant nucleon, and the width of the transverse Gaussian $\sigma_\perp$ is set to 0.5 fm.

Fig.~\ref{fig:hotmap_sub_2D_entropy} shows the entropy density distribution from participating nucleons in the transverse plane for d+Au collisions. Event 0, involving one nucleon from the deuteron and multiple nucleons from the Au nucleus, exhibits a small Gaussian range with light color, indicating an ultra-peripheral collision. In contrast, Event 121, with two nucleons from the deuteron and multiple nucleons from the Au nucleus, shows a broader Gaussian distribution with darker color, corresponding to an ultra-central collision. Events 9, 499, and 786 represent intermediate scenarios between these two extremes. These distinct distributions reflect fluctuations in initial collision conditions.

However, the envelope function alone cannot accurately describe the final-state $\rm dN_{ch}/d\eta$ distribution in d+Au collisions. To address this, we extend Eq.~\eqref{equation:entropy_old} by adding an extra contribution. Physically, the refined model consists of a superposition of three components: wounded nucleons from the deuteron, wounded nucleons from the gold nucleus, and binary collisions. Based on this formulation, we propose a new three-dimensional entropy density distribution as follows:
\begin{equation}\label{entropy_new}
    \begin{aligned}
        \frac{\mathrm{d}S}{\tau_0\mathrm{d}x\mathrm{d}y\mathrm{d}\eta}
        &=g_L f_{L}(\eta_s) \sum_{i=1}^{N_{\mathrm{part}}^{\mathrm{left}}}\rho(\mathbf{x}_\perp, \mathbf{x}_i)\\
         &+ g_R f_R(\eta_s)\sum_{i=1}^{N_{\mathrm{part}}^{\mathrm{right}}}\rho(\mathbf{x}_\perp, \mathbf{x}_i)\\
        & + g_P  f_{plat}(\eta_s)\left[\sum_{i=1}^{N_{\mathrm{part}}^{\mathrm{left}}}\rho(\mathbf{x}_\perp, \mathbf{x}_i)  \sum_{i=1}^{ N_{\mathrm{part}}^{\mathrm{right}}}\rho(\mathbf{x}_\perp, \mathbf{x}_i)\right]^{\beta}
    \end{aligned}    
\end{equation}
The longitudinal entropy density profile is characterized by three distinct deposition functions: $f_{L}(\eta_s)$(left-going), $f_{R}(\eta_s)$(right-going) and $f_{plat}(\eta_s)$(plateau region).  
To properly weight these components, we introduce corresponding coefficients  $g_L$, $g_R$ and $g_P$ subject to the symmetry constraint $g_L = g_R$ that ensures longitudinal symmetry in all parameter configurations. These weighting coefficients will be quantitatively determined through systematic comparison of our model simulations with experimental measurements.
$\beta$ is defined as the entropy deposition coefficient, which significantly affects the contribution of entropy deposition of the interaction term. 
 
The envelope function is defined as follows:
\begin{equation}\label{equation:f_plat}
    f_{\rm plat}(\eta_s) = \exp \left[ -\frac{(|\eta_s| - \eta_s^{\rm plat})^2}{2\sigma_{\eta_{gw}}^2} \theta(|\eta_s| - \eta_s^{\rm plat}) \right],
\end{equation}
where $f_{\mathrm{plat}}(\eta_s)$ characterizes the entropy density distribution in the mid-rapidity region, modeled as a plateau with a width of $\eta_s^{\mathrm{plat}}$. The parameter $\sigma_{\eta_{\mathrm{gw}}}$ governs the Gaussian smearing applied at the edges of the plateau. The Heaviside step function $\theta(|\eta_s| - \eta_s^{\mathrm{plat}})$ ensures that the Gaussian smearing is activated only when $|\eta_s| > \eta_s^{\mathrm{plat}}$.

The left-going component $f_{L}(\eta_s)$ and the right-going component $f_{R}(\eta_s)$ are defined as,
\begin{equation}\label{equation:f_LR}
    f_{L,R}(\eta_s) = \exp\left(-\frac{(\eta_s \pm (\eta_s^{0} - \Delta \eta_s^{L,R}))^2}{2 \sigma^2}\right),
\end{equation}
where the $+(-)$ sign corresponds to L (R). Here two Gaussian profiles centered at $\mp(\eta_s^{0} - \Delta \eta_s^{L,R})$ are employed, $\eta_s^{0}=5.36$ is the beam rapidity in the center-of-mass frame of two colliding nucleons and $\Delta \eta_s^{L,R} > 0$ quantifies the pseudo-rapidity shift due to rapidity loss in the deuteron and gold nuclei. These components share a common width parameter $\sigma$ that determines their Gaussian fall-off.

The large asymmetry in nucleon numbers (2 vs. 197) leads to distinct collision dynamics. Gold nucleons experience at most one or two collisions regardless of centrality, while deuteron nucleons collide more frequently, especially in central events. Consequently, the rapidity loss $\Delta\eta_s^R$ of the deuteron exhibits strong centrality dependence, whereas that of the gold nucleus, $\Delta\eta_s^L$, is much smaller than $\Delta\eta_s^R$. This asymmetric treatment effectively captures the essential features of the initial entropy density distribution.

In our baseline model, we simply take constant pseudo-rapidity losses $\Delta\eta_s^L = 1.36$ and $\Delta\eta_s^R = 4.36$ for all centralities. To account for the enhanced energy loss of deuteron nucleons in central collisions due to multiple scatterings, we introduce a centrality-dependent parametrization only for $\Delta\eta_s^R$:
\begin{equation}\label{eq:delta_eta}
    \Delta \eta_s^{R}(n_{\rm BC})
    =\frac{n_{\rm BC}-n_{\rm BC}^{\rm {min}}}{n_{\rm BC}^{\rm {max}}-n_{\rm BC}^{\rm {min}}} + \Delta\eta_s^{R}
\end{equation}
Here $n_{\rm BC}$ counts the number of binary nucleon-nucleon collisions in a single event, while $n_{\rm BC}^{\rm min}$ and $n_{\rm BC}^{\rm max}$ denote the minimum and maximum values observed across all events, respectively. $\Delta\eta_s^L$ remains fixed at 1.36.

The entropy deposition coefficient $\beta$ characterizes the degree of local entropy density deposition in the interaction term of d+Au collisions. It exponentially scales the accumulated entropy value at each grid point on the transverse plane, which is closely related to the formation of local entropy density in the central rapidity region during collisions. This assumption stems from the nucleus-nucleus collision framework, where we describe the energy flux density through the following expression:
\begin{equation}
    \mathbf{J}_{A}^\mu = T_{A} (E_{A}, 0, 0, p_{A}),\,\,
    \mathbf{J}_{B}^\mu = T_{B} (E_{B}, 0, 0, p_{B}) 
\end{equation}
When $E \gg m $,  in the  center-of-mass frame, $E_A \approx |p_A| = |p_B| \approx E_B$. Defining $|p_A| = |p_B| \equiv p$, we have
\begin{equation}
    \mathbf{J}_{A}^\mu = T_{A} (p, 0, 0, p) ,\,\,
    \mathbf{J}_{B}^\mu = T_{B} (p, 0, 0, -p) 
\end{equation}
For a pair of nucleons colliding in the center-of-mass frame, $p_1^{\mu} = (p, 0, 0, p) $, $p_2^{\mu} = (p, 0, 0, -p) $, so $s_\mathrm{{NN}}=(p_1 +p_2)^{2}=4p^{2} $, so we can get the transverse-area density of the center-of-mass energy is
\begin{equation}\label{fireball_beta}
    \begin{aligned}
        \frac{dM}{d^2 \mathbf{x}_\perp} = \sqrt{(\mathbf{J}_A^\mu + \mathbf{J}_B^\mu)^2} 
         = (T_A \cdot T_B)^{1/2} \, \sqrt{s_{\rm NN}}\,.
    \end{aligned}
\end{equation}
Therefore, the total initial energy deposition is expected to locally scale as $(T_A T_B)^{1/2}$.

At early times, when the longitudinal expansion dominates over microscopic scattering, the system approximately undergoes free-streaming Bjorken expansion. As the longitudinal pressure becomes small, both the energy density and the kinetic entropy density dilute approximately as $1/\tau$. Provided that the ratio between the deposited energy and the produced coarse-grained entropy is approximately independent of transverse position, the entropy density at hydrodynamization inherits the same local thickness-function dependence.
This motivates $\beta=0.5$ as a representative exponent characterizing entropy deposition in the nuclear overlap term, a choice that has also been widely adopted in initial-condition models. In this study, we systematically investigate how variations of $\beta$ affect both the initial entropy-density distribution and the final-state hadron distributions in d+Au collisions.

In Table~\ref{tab:parameter_for_IC_3D}, we present four distinct parameter sets (a), (b), (c), and (d) for generating initial entropy density profiles. These parameter configurations enable the production of corresponding initial entropy density distributions, which are subsequently employed in relativistic hydrodynamic simulations to investigate both the impact of different physical assumptions on initial conditions and their effects on final-state charged particle multiplicity distributions.

\begin{table*}[htp]
\caption{\label{tab:parameter_for_IC_3D}
Four parameter sets for 3D initial condition.}
\renewcommand{\arraystretch}{1.2} 
\begin{ruledtabular}
\begin{tabular}{cccccccccccc}
&\textrm{Parameter Sets}
&\textrm{$\eta_s^{\text{plat}}$}
&\textrm{$\sigma_{\eta_{\text{gw}}}$}
&\textrm{$\eta_s^{0}$}
&\textrm{$\sigma$}
&\textrm{$ g_{L}$}
&\textrm{$ g_{R}$}
&\textrm{$ g_{P}$}
&\textrm{$\beta$}
&\textrm{$\Delta \eta_s^L$(Au)}
&\textrm{$\Delta \eta_s^R$(d)}\\
\colrule
& (a) & 1.3 & 1.3 & 5.36 & 2.5 & 8.0 & 8.0 & 22.5  & 0.5  & 1.36 & 4.36  \\
& (b) & $\uparrow$ & $\uparrow$ & $\uparrow$ & $\uparrow$ & $\uparrow$ & $\uparrow$ & $\uparrow$  & 0.4  & 1.36 & 4.36\\ 
& (c) & $\uparrow$ & $\uparrow$ & $\uparrow$ & $\uparrow$ & $\uparrow$ & $\uparrow$ & $\uparrow$  & 0.35 & 1.36 & 4.36\\ 
& (d) & $\uparrow$ & $\uparrow$ & $\uparrow$ & $\uparrow$ & $\uparrow$ & $\uparrow$ & $\uparrow$  & 0.35 & 1.36 & $\Delta \eta_s(n_{\rm BC})$\\ 
\end{tabular}
\end{ruledtabular}
\begin{flushleft}
\small \textit{\rm Note:} The symbol $\uparrow$ indicates repetition of the value from the preceding row. For set (d), $\Delta \eta_s(n_{\rm BC})$ is a function of $n_{\rm BC}$ defined as Eq.~\eqref{eq:delta_eta}.
\end{flushleft}
\end{table*}

In the numerical implementation, the spacetime rapidity interval $[-\eta_{s,{\rm max}}, +\eta_{s,{\rm max}}] = [-6.9, 6.9]$ is uniformly discretized into 121 grid points. For the transverse plane, we adopt a square grid in $x$ and $y$ with different ranges depending on the collision system. For d+Au collisions, the grid covers $[-10, +10]$~fm in both dimensions and is discretized into $200 \times 200$ cells, yielding a spatial resolution of $0.1$~fm per cell. For Au+Au collisions, the grid spans $[-16, +16]$~fm, also using $200 \times 200$ cells, which gives a resolution of $0.16$~fm per cell. This setup produces a three-dimensional entropy density tensor of size $(200, 200, 121)$ corresponding to the $(x, y, \eta_s)$ coordinates.

\subsection{Determining Centrality from Initial Conditions}\label{subsec:centrality-determination}

Accurate centrality determination is essential for meaningful comparisons between theoretical calculations and experimental measurements. In the PHOBOS experiment for d+Au collisions, centrality is determined based on the energy deposition measured by ring counters within the pseudorapidity range $3.0 < |\eta| < 5.4$, which is proved to be proportional to the number of charged multiplicity, in the experiment~\cite{PHOBOS:2004fzb}.  
Following the experimental procedure, a straightforward approach would involve performing full hydrodynamic simulations for all events and classifying them based on the final-state charged particle multiplicity within the same pseudorapidity range.  
However, conducting thousands to tens of thousands of event-by-event hydrodynamic simulations for centrality determination is computationally intensive, even though the single-event hydrodynamic code CLVisc achieves high efficiency through GPU parallelization.  
To circumvent this substantial computational cost, we adopt a centrality classification method based on the event-by-event longitudinal entropy distribution in the initial state. Later, we will demonstrate that the integrated entropy within the same space-time rapidity range $3.0 < |\eta_s| < 5.4$ provides a good approximation to the event ordering based on charged particle multiplicity.

The longitudinal entropy distribution as a function of space-time rapidity $\eta_s$ is defined as:
\begin{equation}\label{equation:dsdeta_dis}
    \frac{\mathrm{d}S}{\mathrm{d}\eta_s}=\int \frac{\mathrm{d}S}{\tau_0\mathrm{d}x\mathrm{d}y\mathrm{d}\eta_s} \tau_0 \mathrm{d}x\mathrm{d}y .
\end{equation}
Using Eq.~\eqref{equation:dsdeta_dis}, we compute $\mathrm{d}S/\mathrm{d}\eta_s$ for each of the 50,000 initial condition events using our longitudinal entropy deposition model. To emulate the experimental centrality determination in the PHOBOS measurement, we integrate $\mathrm{d}S/\mathrm{d}\eta_s$ over the same range $3.0 < |\eta_s| < 5.4$ to obtain the total entropy $dS$ within this region. Centrality intervals are then defined by sorting all events in descending order of their entropy values.

\begin{figure*}
    \centering
    \includegraphics[width=0.85\textwidth]{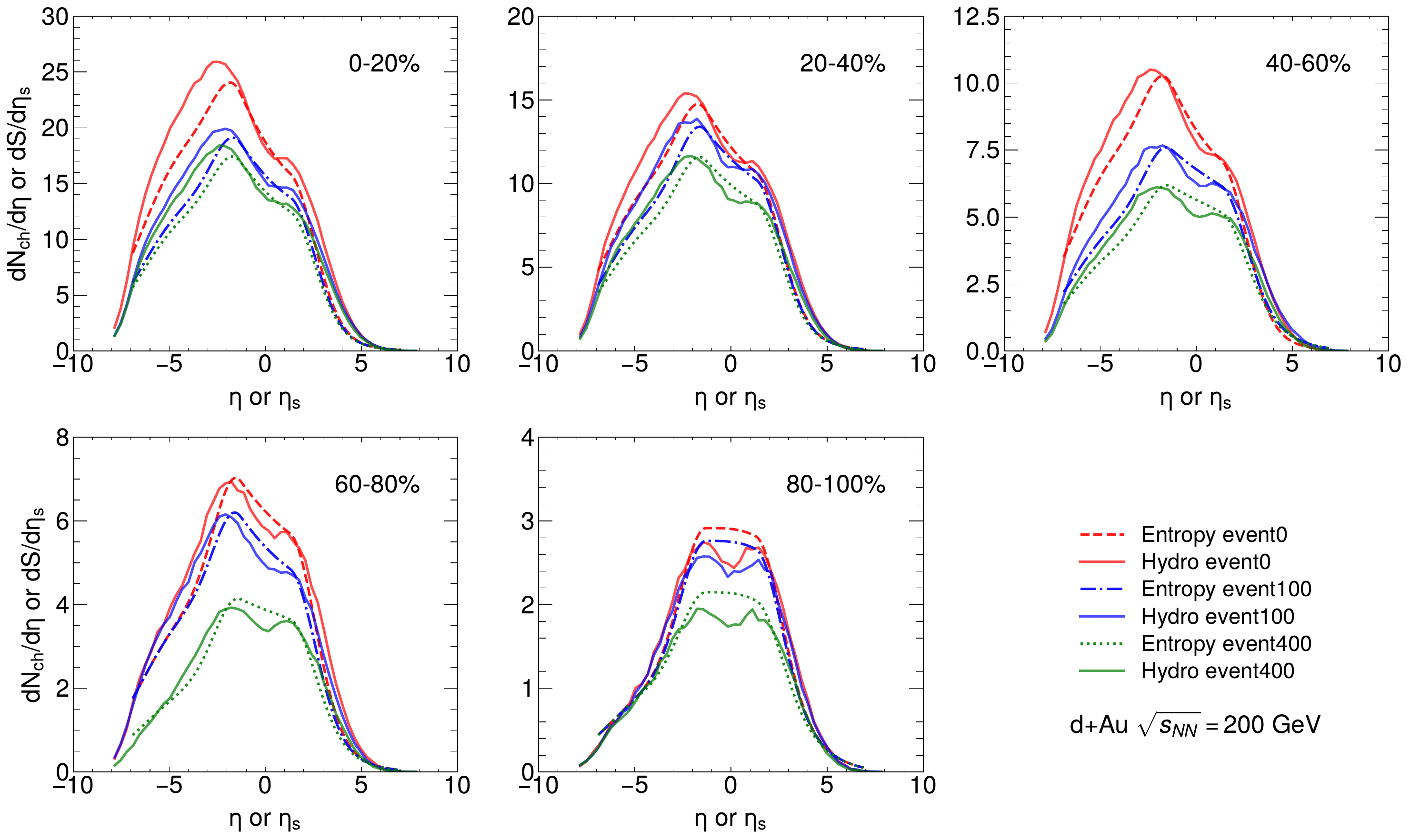} 
    \caption {(Color online) Distributions of $\mathrm{d}N_{\mathrm{ch}}/\mathrm{d}\eta$ and $\mathrm{d}S/\mathrm{d}\eta_s$ for d+Au collisions in five centrality classes ($0\%$--$20\%$, $20\%$--$40\%$, $40\%$--$60\%$, $60\%$--$80\%$, and $80\%$--$100\%$) using parameter set (a) in Table~\ref{tab:parameter_for_IC_3D}. For each centrality bin, three representative events are shown. Solid lines: $\mathrm{d}N_{\mathrm{ch}}/\mathrm{d}\eta$ from CLVisc; dotted lines: corresponding $\mathrm{d}S/\mathrm{d}\eta_s$ distributions. The relative ordering of $\mathrm{d}S/\mathrm{d}\eta_s$ among events remains unchanged after hydrodynamic evolution, and the same ordering is preserved for $\mathrm{d}N_{\mathrm{ch}}/\mathrm{d}\eta$, demonstrating that the centrality classification scheme based on $\mathrm{d}S$ is consistent and feasible. }
    \label{fig:dNdeta_VS_dSdeta}
\end{figure*}

\begin{figure}
    \centering
    \includegraphics[width=0.47\textwidth]{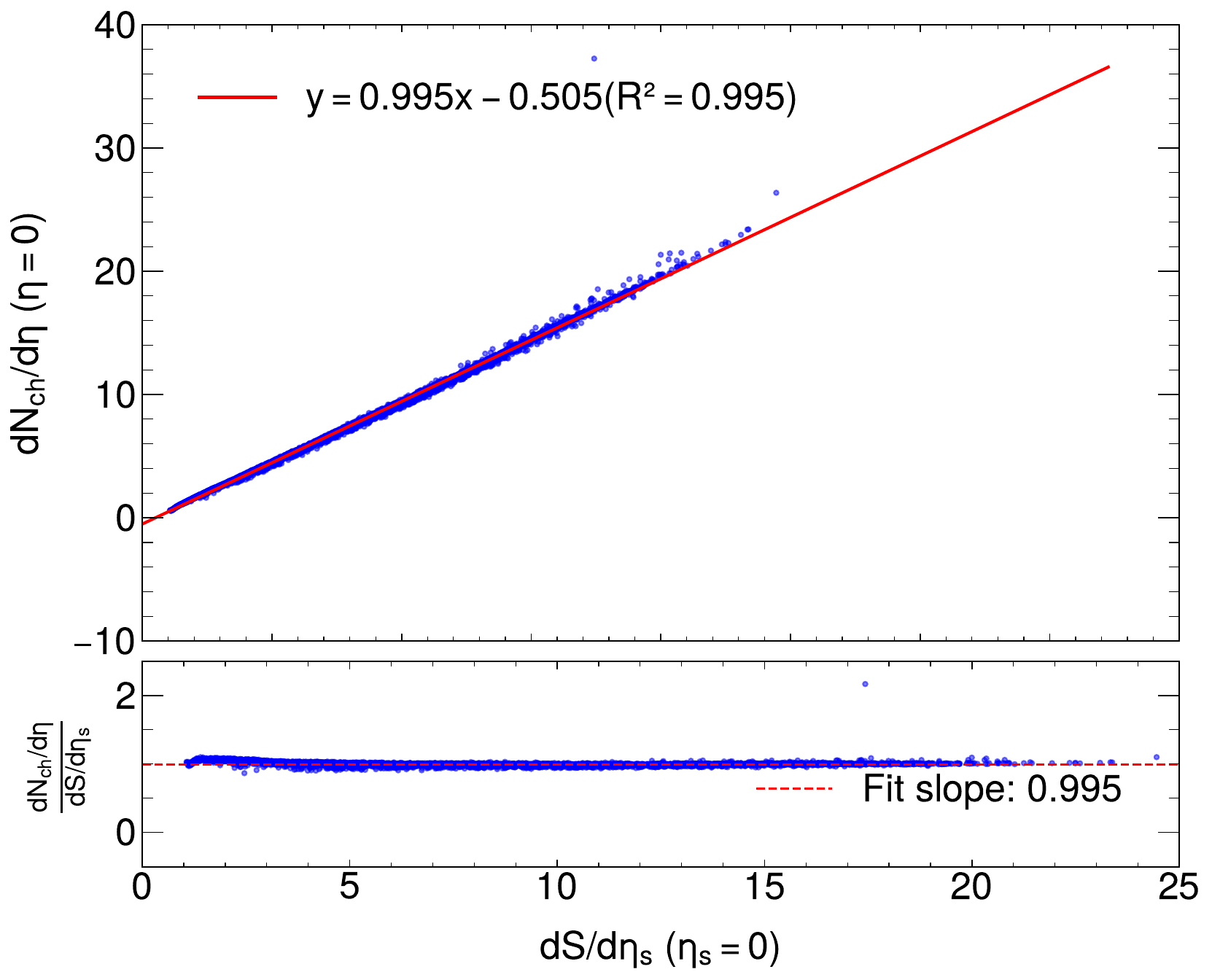}
    \caption{(Color online)Upper panel: two-dimensional scatter plot (solid blue dots) of the final-state charged particle multiplicity $\mathrm{d}N_{\mathrm{ch}}/\mathrm{d}\eta$ versus the initial-state longitudinal entropy density distribution $\mathrm{d}S/\mathrm{d}\eta_s$ at $\eta = 0$ and $\eta_s = 0$, computed using parameter set (a) in Table~\ref{tab:parameter_for_IC_3D} for 5000 events. The solid red line shows the linear fit to the scatter data. Lower panel: ratio $\frac{\mathrm{d}N_{\mathrm{ch}}/\mathrm{d}\eta}{\mathrm{d}S/\mathrm{d}\eta_s}$ as a function of $\mathrm{d}S/\mathrm{d}\eta_s$, where the numerator is shifted by subtracting the intercept ($-0.505$) obtained from the linear fit. }
    \label{fig:dNdeta_dSdeta_ratio}
\end{figure}

\begin{figure*}
    \centering
    \includegraphics[width=0.85\textwidth]{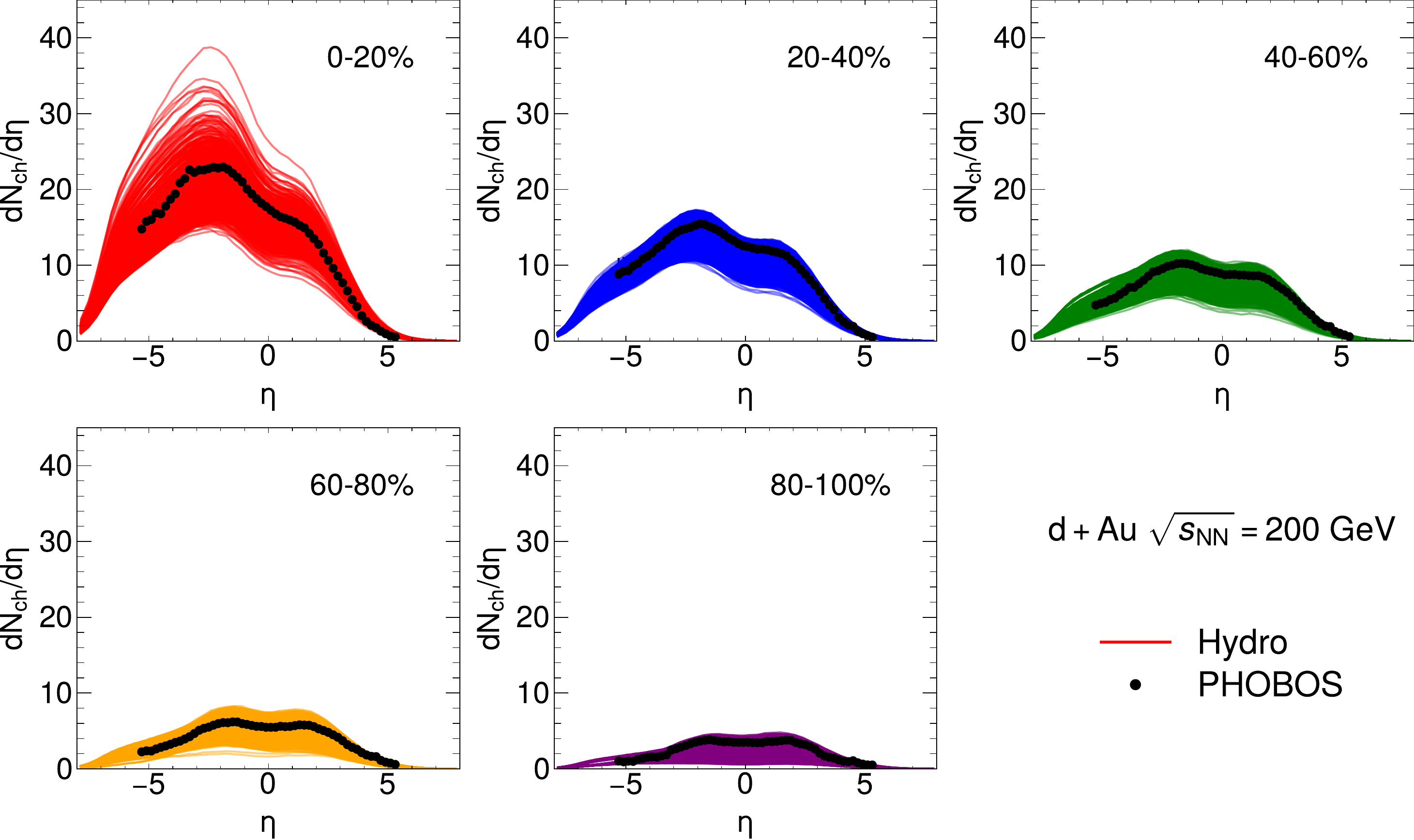}
    \caption {(Color online) Pseudorapidity distributions $\mathrm{d}N_{\mathrm{ch}}/\mathrm{d}\eta$ for d+Au collisions in five centrality classes ($0\%$--$20\%$, $20\%$--$40\%$, $40\%$--$60\%$, $60\%$--$80\%$, and $80\%$--$100\%$) using parameter set (a) in Table~\ref{tab:parameter_for_IC_3D}. Colored solid lines: event-by-event hydrodynamic simulations from CLVisc; black scattered dots: experimental data from the PHOBOS collaboration~\cite{PHOBOS:2004fzb}. }
    \label{fig:dNdeta_ebe}
\end{figure*}

Fig.~\ref{fig:dNdeta_VS_dSdeta} compares the $\rm dN_{ch}/d\eta$ distributions obtained from event-by-event hydrodynamic simulations with the initial $\mathrm{d}S/\mathrm{d}\eta_s$ distributions. The initial $\mathrm{d}S/\mathrm{d}\eta_s$ distributions maintain the same relative ordering as the final-state $\rm dN_{ch}/d\eta$ distributions after hydrodynamic evolution. Specifically, events with higher initial entropy density yield correspondingly higher charged particle multiplicities, and vice versa. This one-to-one correspondence validates the feasibility of our current centrality determination method.

In central collisions, the final-state $\rm dN_{ch}/d\eta$ distribution obtained from hydrodynamic simulations numerically exceeds the initial $\mathrm{d}S/\mathrm{d}\eta_s$ distribution. However, as the collisions become more peripheral, the charged multiplicity result begins to fall below the initial entropy density distribution. This may be because, in peripheral d+Au collisions, the temperature of the produced QGP is relatively low, and due to the shorter evolution time and the smaller entropy production during evolution (caused by non-zero shear viscosity), resulting in a lower final-state particle yield. 

Fig.~\ref{fig:dNdeta_dSdeta_ratio} displays a scatter plot comparing the final-state charged-particle multiplicity density $\rm dN_{ch}/d\eta$ at $\eta=0$ with the initial-state entropy density $\mathrm{d}S/\mathrm{d}\eta_s$ at $\eta_s = 0$ for individual hydrodynamic events. A strong linear correlation further supports our use of initial-state $\mathrm{d}S/\mathrm{d}\eta_s$ for centrality classification. It is noteworthy that the linear fit yields a non-zero intercept, implying that the relationship between $\mathrm{d}S/\mathrm{d}\eta_s$ and $\rm dN_{ch}/d\eta$ is not purely proportional.

Fig.~\ref{fig:dNdeta_ebe} presents the $\rm dN_{ch}/d\eta$ distributions obtained from the event-by-event hydrodynamic simulations and compares them with the PHOBOS experimental data (event-averaged result). The observed large fluctuations originate from initial-state geometry fluctuations—including the distribution of nucleons within the nucleus, the random orientation of the colliding nuclei, and a centrality-dependent rapidity loss—as well as from Monte Carlo sampling during the final-state hadronization stage. The event-averaged $\rm dN_{ch}/d\eta$ from PHOBOS lies on top of our event-by-event simulation results. Our model reproduces the overall trend of the PHOBOS data and the centrality determination.

Different centrality classes correspond to distinct initial-state geometric eccentricities, defined as:
\begin{equation}
    \varepsilon_n e^{i n \phi_n} = -\frac{\int dx\, dy\, r^n e^{i n \phi} S_{xy}}{\int dx\, dy\, r^n S_{xy}},
\end{equation}
Here, $\varepsilon_n$ is the $n$-th order geometric eccentricity, characterizing the spatial anisotropy of the initial entropy density distribution, and $\phi_n$ denotes the $n$-th order participant plane. The weight function $S_{xy} \equiv \left. \frac{dS}{dx dy d\eta_s} \right|_{\eta_s=0}$ represents the local entropy density in the transverse plane at midrapidity, where the majority of produced particles are observed. These eccentricities are used to investigate the influence of nuclear structure effects on the initial state of d+Au collisions at $\sqrt{s_{\mathrm{NN}}} = 200\ \text{GeV}$.

\subsection{Hydrodynamics Simulation and Afterburner}

Relativistic hydrodynamics provides an effective description of the QGP spacetime evolution in high-energy heavy-ion collisions. The conservation equations for the energy-momentum tensor $T^{\mu\nu}$ and net baryon current $N^\mu$ are given by:
\begin{equation}
    \nabla_\mu T^{\mu\nu} = 0,\quad \nabla_\mu N^\mu = 0,
\end{equation}
where $\nabla_\mu$ denotes the covariant derivative. In the Landau frame, $T^{\mu\nu}$ is decomposed as:
\begin{equation}
    T^{\mu\nu} = \epsilon u^\mu u^\nu - (p + \Pi)\Delta^{\mu\nu} + \pi^{\mu\nu},
\end{equation}
with $\epsilon$ the energy density, $p$ the pressure, $u^\mu$ the fluid four-velocity normalized as $u_\mu u^\mu = 1$, and $\Delta^{\mu\nu} = g^{\mu\nu} - u^\mu u^\nu$ the projection operator orthogonal to the fluid velocity. The bulk pressure $\Pi$ and shear stress tensor $\pi^{\mu\nu}$ account for dissipative corrections. The shear tensor is traceless, $\pi^\mu_\mu = 0$, and transverse to the fluid velocity, $u_\mu \pi^{\mu\nu} = 0$. Their evolution follows the Israel-Stewart equations, ensuring causal propagation of dissipative effects.

We employ the (3+1)-dimensional hydrodynamic model CLVisc to simulate the QGP evolution and compute the $dN_{ch}/d\eta$, $p_T$ spectra, and anisotropic flows in d+Au and Au+Au collisions at $\sqrt{s_{\rm NN}}=200$ GeV. The initial conditions are obtained from the asymmetric rapidity loss model described in Sec.~\ref{subsec:3d-initial-condition} and \ref{subsec:centrality-determination}, which provides the initial entropy density distribution. Bulk viscosity and net baryon charge conservation are neglected.

Neglect of net baryon charge conservation is well justified at $\sqrt{s_{\rm NN}} = 200$~GeV, where the baryon chemical potential $\mu_B \approx 20$--$30$ MeV is small compared to the temperature scale~\cite{Andronic:2005yp,Andronic:2008gu}. Bulk viscosity is also neglected in the present study. Although the bulk viscosity to entropy density ratio $\zeta/s$ is known to peak near the QCD crossover temperature~\cite{Karsch:2007jc,Meyer:2007dy} and can play a non-negligible role in certain observables~\cite{Ryu:2017qzn,Gardim:2020mmy}, a recent Bayesian analysis of RHIC Beam Energy Scan data finds that the specific bulk viscosity exhibits a preferred maximum around $\sqrt{s_{\rm NN}} = 19.6$~GeV and decreases towards top RHIC energies~\cite{Shen:2023awv}. Therefore, its effect on the validation of our model at $\sqrt{s_{\rm NN}} = 200$~GeV is expected to be minor. Since the goal of this work is not a precision extraction of transport coefficients, the neglect of bulk viscosity is a reasonable simplification for the present purpose. A systematic study including bulk viscosity is left for future investigations.

After the hydrodynamic expansion, the system is converted into hadrons on a freeze-out hypersurface via the Cooper-Frye prescription. To account for the subsequent hadronic cascade, we employ the SMASH afterburner~\cite{SMASH:2016zqf}, which handles hadron propagation, scattering, resonance formation and decays. For d+Au collisions, the freeze-out temperature is set at $T_{\mathrm{frz}} = 0.150$ GeV. 

For Au+Au collisions, to save computational resources, we run hydrodynamic simulation till $T_{\mathrm{frz}} = 0.128$ GeV, which spans the evolution region between chemical freeze-out and kinetic freezeout. The sampled hadrons are directly converted into stable particles through resonance decay.
We have verified that the SMASH after-burner does not yield a visible difference for $dN_{ch}/d\eta$ in d+Au systems. Consequently, in the subsequent validation of charged rapidity distributions of p+Au and $^3$He+Au, hydrodynamic simulations are likewise extended down to kinetic freeze-out.

For the $p_T$ spectra calculations, we follow the experimental protocol. In both d+Au and Au+Au collisions, a pseudorapidity cut of $|\eta| < 0.35$ is applied to select charged particles. The identified particle spectra $d^2N/(2\pi p_T dp_T dY)$ are then computed with rapidity $Y$. 

In Au+Au collisions, the anisotropic flow coefficients are obtained using the $Q_n$ vector method,
\begin{equation}\label{eq:Qn}
    Q_n = \frac{1}{M} \sum_{i=1}^{M} e^{in\phi_i},
\end{equation}
where $M$ is the number of particles in a single event and $\phi_i$ the azimuthal angle of the $i$-th particle. In our simulations, we sample the freeze-out hypersurface 2000 times per event to ensure a sufficiently large $M$. In the large-$M$ limit, the single-event average becomes:
\begin{align}
    \widetilde{Q}_n &= \lim_{M \rightarrow \infty} Q_n = \left\langle e^{in\phi} \right\rangle_{\text{particles}} = v_n^{\text{(ev)}} e^{i n \Psi_n},
\end{align}
with $v_n^{\text{(ev)}}$ the event-wise flow coefficient and $\Psi_n$ the event plane angle. The event-averaged flow coefficient is then:
\begin{equation}
    v_n = \left\langle v_n^{\text{(ev)}} \right\rangle_{\text{events}} = \left\langle |\widetilde{Q}_n| \right\rangle_{\text{events}}.
\end{equation}

For d+Au collisions, we follow the PHENIX event plane method:
\begin{equation}
    v_n^{\{\text{EP}\}} \equiv \frac{\left\langle \cos\left[ n\left( \phi - \Psi_n^{\text{Ref.}} \right) \right] \right\rangle}{R\left( \Psi_n^{\text{Ref.}} \right)},
\end{equation}
where $\Psi_n^{\text{Ref}}$ is the $n$th-order reference event plane and $R$ its resolution, estimated from subevent correlations:
\begin{equation}
    R\left( \Psi_n^{\text{Ref.}} \right) = 
    \sqrt{\frac{\left\langle\frac{Q_{nA}}{|Q_{nA}|}\frac{Q_{nB}^*}{|Q_{nB}|}\right\rangle_\text{events} \left\langle\frac{Q_{nA}}{|Q_{nA}|}\frac{Q_{nC}^*}{|Q_{nC}|}\right\rangle_\text{events}}{\left\langle\frac{Q_{nB}}{|Q_{nB}|}\frac{Q_{nC}^*}{|Q_{nC}|}\right\rangle_\text{events}}}.
\end{equation}
The flow vectors $Q_{nA}$, $Q_{nB}$, and $Q_{nC}$ are defined in the pseudorapidity regions $-3.9 < \eta < -3.1$ (south BBC), $-3.0 < \eta < -1.0$ (south FVTX), and $|\eta| < 0.35$ (central CNT), respectively~\cite{PHENIX:2018lia}.

In this study, for both systems, the computational grid is set to $n_{\rm x} \times n_{\rm y} \times n_{\rm z} = 200 \times 200 \times 121$, the thermalization time is $\tau_0 = 0.6$ fm, the shear viscosity to entropy density ratio is $\eta/s = 0.08$, and the equation of state $lattice-pce165$ is employed. For each configuration and centrality, we generate 1000 hydrodynamic events.

\section{Results}\label{sec:results}

\subsection{Effect of Deuteron Nuclear Structure at initial stage}

\begin{figure*}[htp]
    \includegraphics[width=0.45 \textwidth]{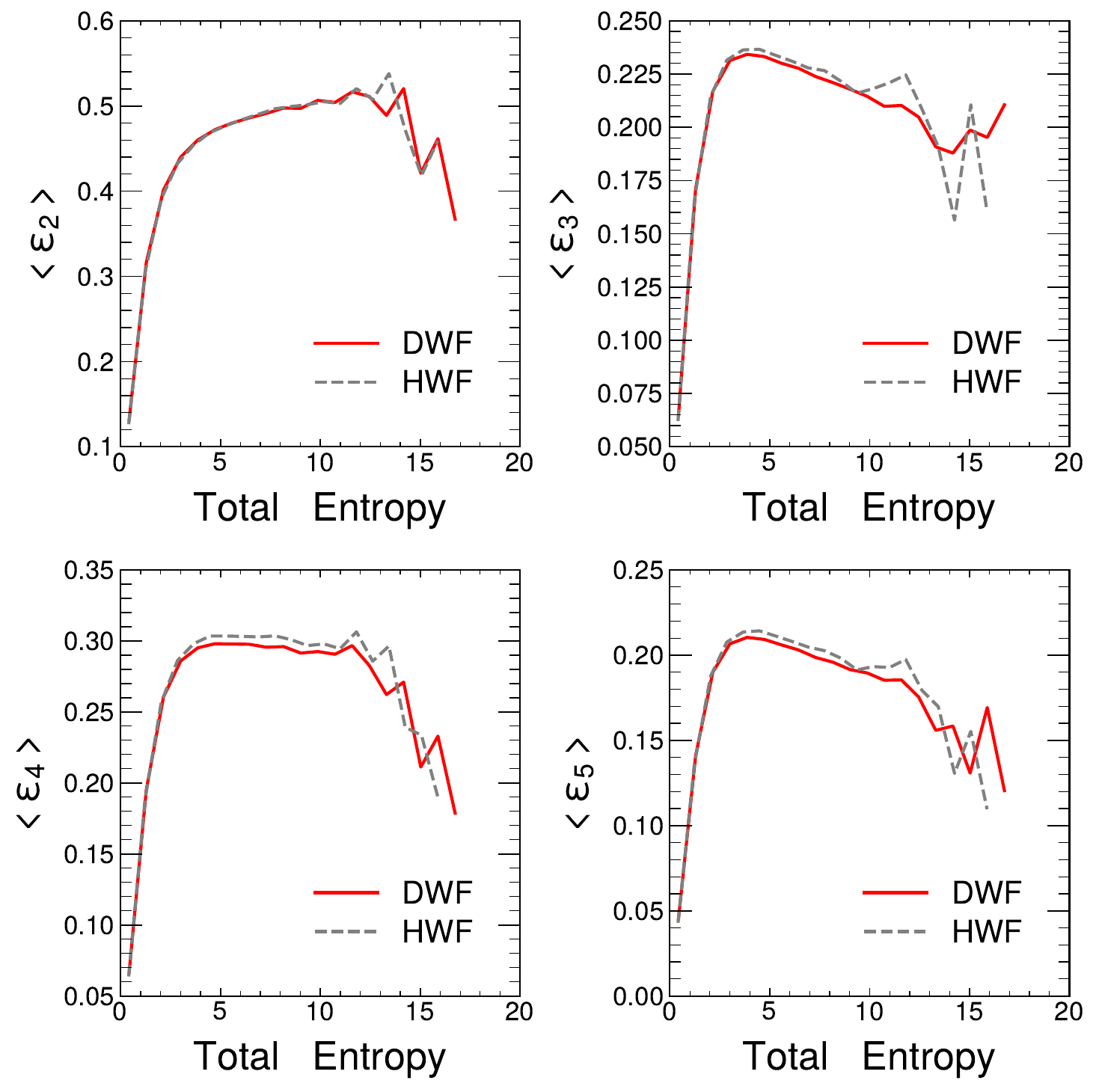}  
    \includegraphics[width=0.45 \textwidth]{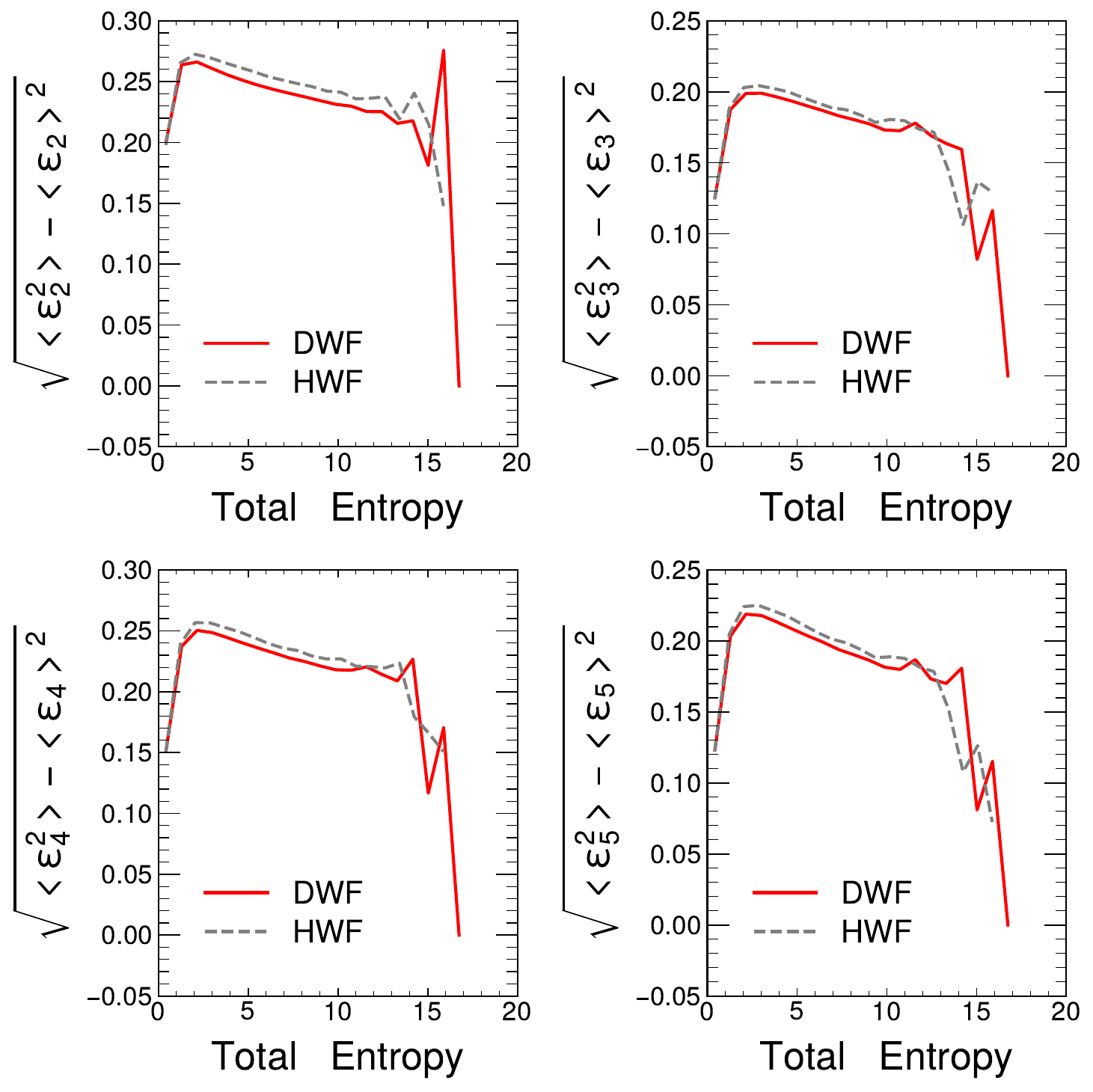} 
    \caption{(Color online) Left four panels: $\langle \varepsilon_n \rangle$ as a function of total entropy; right four panels: fluctuations $\sqrt{\langle \varepsilon_n^2 \rangle - \langle \varepsilon_n \rangle^2}$. Results from the DWF nuclear structure are consistently lower than those from the HWF, reflecting the more compact nucleon configuration in the DWF, which reduces fluctuations in $\langle \varepsilon_n \rangle$. The increase in fluctuations at large total entropy is due to limited event statistics. Solid red lines denote DWF results, and dashed gray lines denote HWF results. }
    \label{fig:epsilon_n_vs_entropy}
\end{figure*}

To test the effects of a more realistic deuteron wave function (DWF) incorporating $Y_{0,0}$–$Y_{2,0}$ interference on anisotropic flows, we compare the initial-state geometric eccentricity $\varepsilon_n$ in \trento-2D for the DWF and the HWF (without interference) initial conditions.
Note that \trento-2D is used only in Fig.~\ref{fig:epsilon_n_vs_entropy} in this paper, which shows the mean (left four panels) and fluctuations (right four panels) of $\varepsilon_n$ as functions of total entropy for $n = 2, 3, 4, 5$.
The mean value $\langle \varepsilon_2 \rangle$ exhibits negligible difference between the DWF and HWF configurations. For $n = 3, 4, 5$, both the mean $\varepsilon_n$ and the fluctuations show only tiny differences between these two initial-condition sets. Such minor differences may be washed out during complex hydrodynamic evolution. The DWF results are systematically smaller than those from HWF, consistent with the denser nucleon distribution of the DWF configuration shown in Fig.~\ref{fig:r_theta_phi_dis_in_DWF_and_HWF}. This demonstrates that the nucleon distribution information is retained in the initial state of deuteron collisions at high energy.

\begin{figure}
    \centering
    \includegraphics[width=0.45\textwidth]{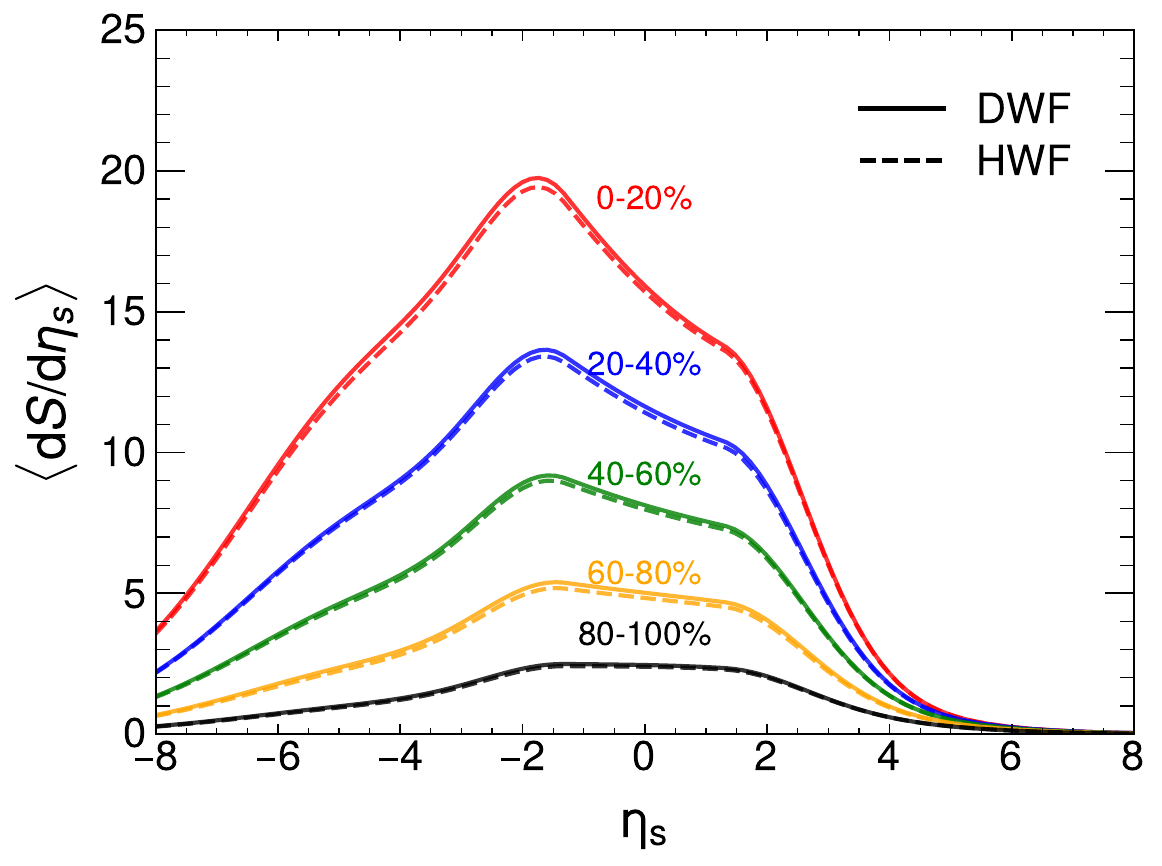}
    \caption {(Color online) Distribution of average $\rm dS/d\eta_s$ of d+Au collision with different centrality $0\%-20\%$, $20\%-40\%$, $40\%-60\%$, $60\%-80\%$, and $80\%-100\%$ of parameters set (a) in Table~\ref{tab:parameter_for_IC_3D}. The colored solid lines represent the results of the DWF structure, and the colored dashed lines represent the results of the HWF structure. }
    \label{fig:Average_dSdeta}
\end{figure}

Fig.~\ref{fig:Average_dSdeta} compares the average $\mathrm{d}S/\mathrm{d}\eta_s$ distributions for DWF and HWF across different centrality classes. Although the nuclear structure differences are evident in the initial state (Fig.~\ref{fig:epsilon_n_vs_entropy}), they become indistinguishable in the longitudinal entropy distribution. Given the strong correlation between $\mathrm{d}S/\mathrm{d}\eta_s$ and the final-state $dN_{ch}/d\eta$, this suggests that the conventional $dN_{ch}/d\eta$ observable is also insensitive to nuclear structure. Therefore, using the more realistic DWF structure provides only marginal improvement in describing the charged particle multiplicity in d+Au collisions. 

\subsection{$\rm dN_{ch}/d \eta$ Distribution with Different Longitudinal Entropy Depositions}

\begin{figure*}[htp]
  \vspace{0.5em}
  \includegraphics[width=0.45 \textwidth]{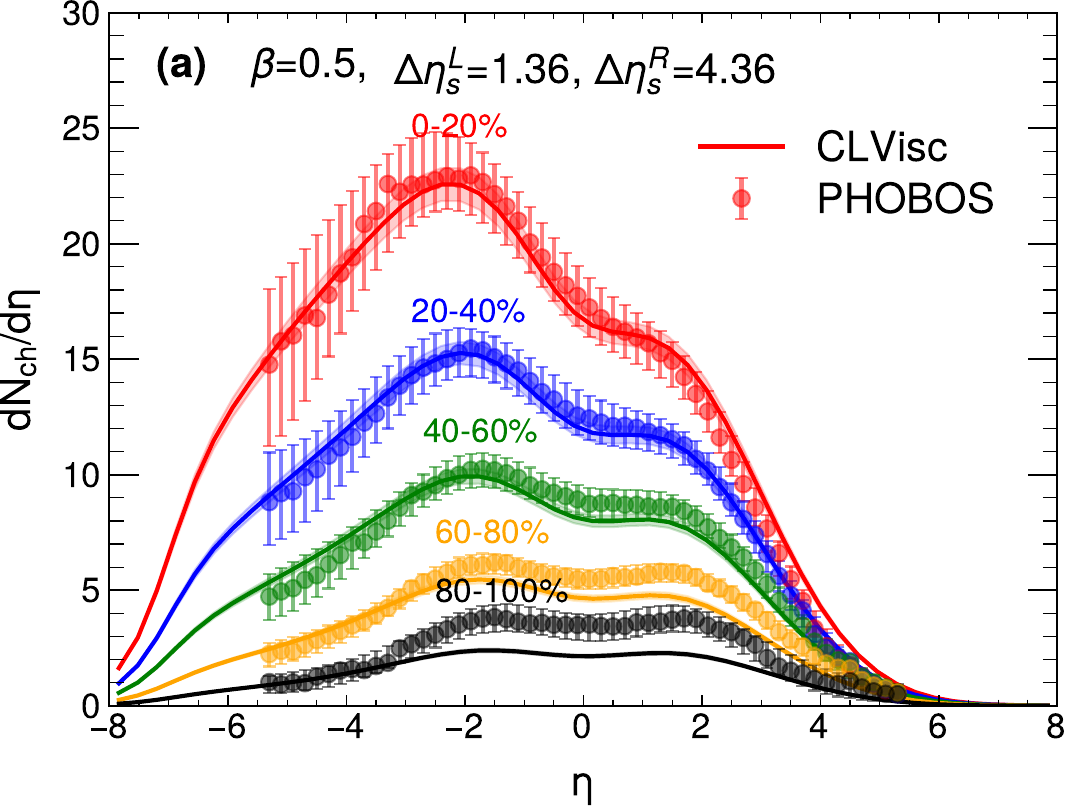}
  \includegraphics[width=0.45 \textwidth]{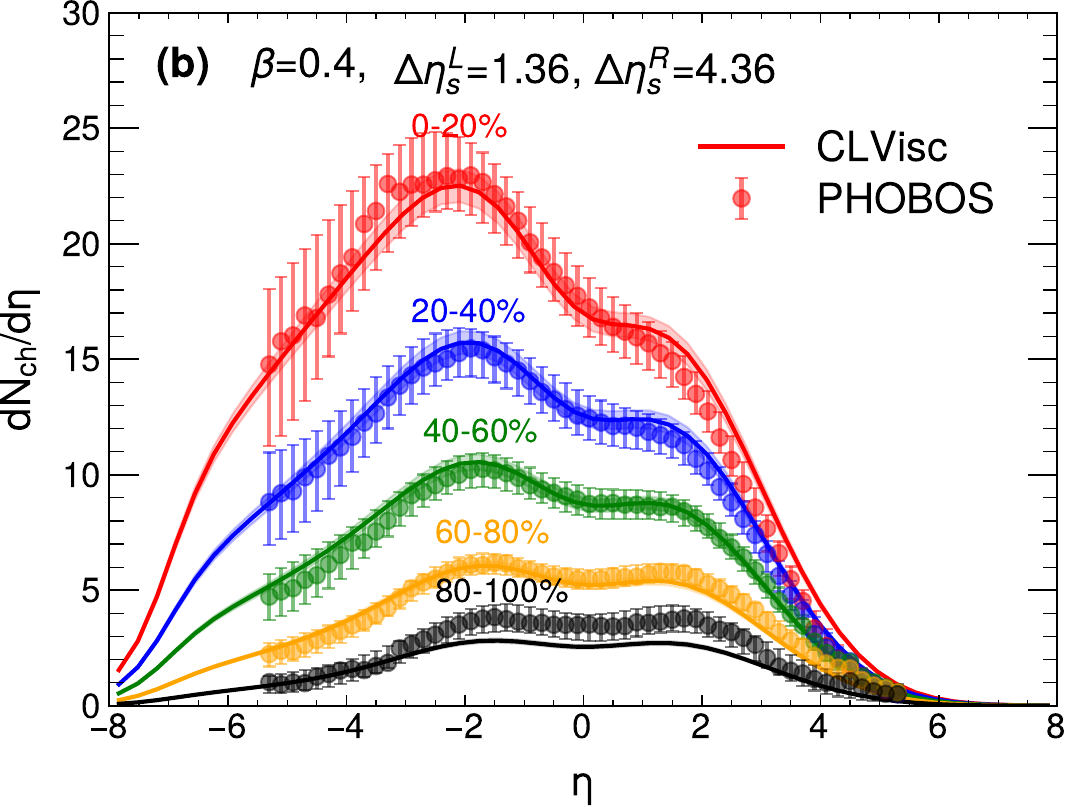} \\
  \includegraphics[width=0.45 \textwidth]{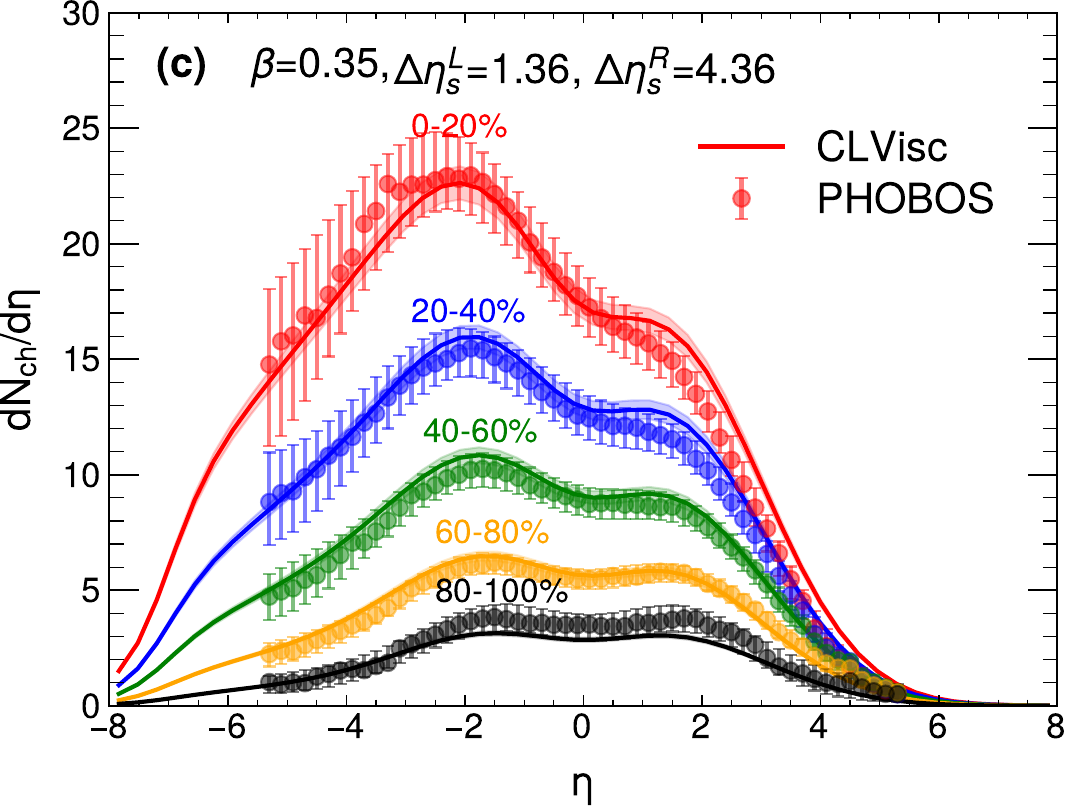}
  \includegraphics[width=0.45 \textwidth]{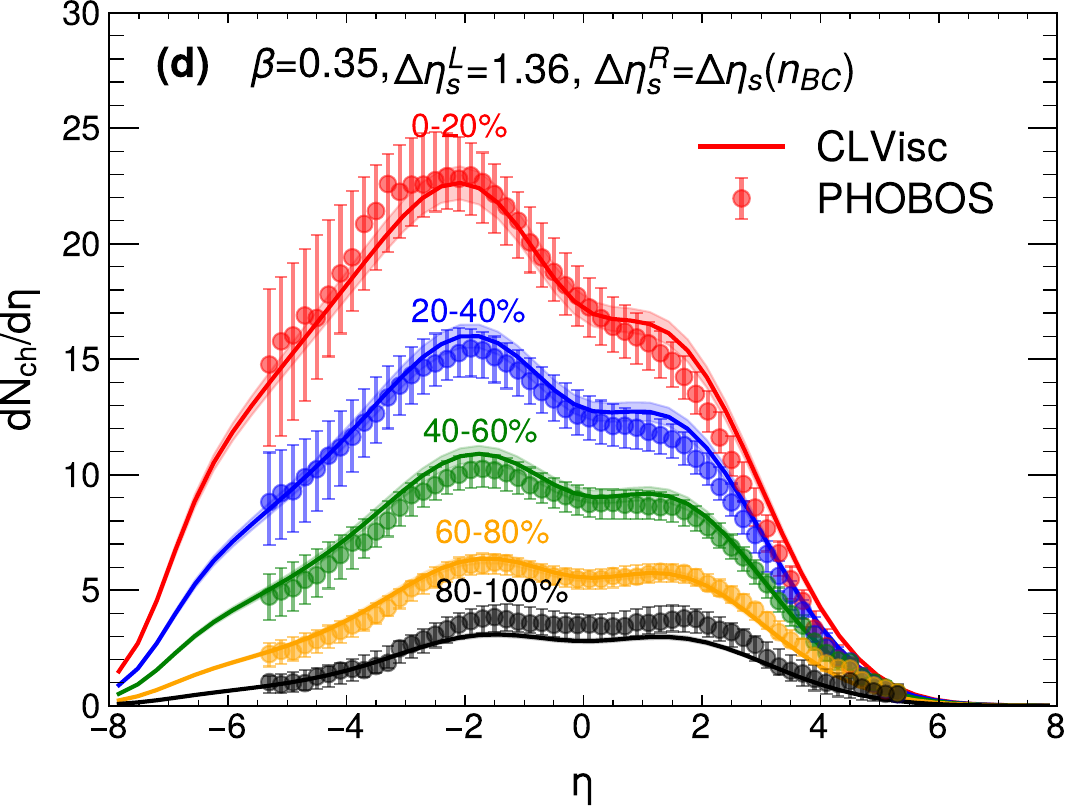}\\
  \caption{(Color online) Pseudorapidity distributions $\mathrm{d}N_{\mathrm{ch}}/\mathrm{d}\eta$ for d+Au collisions at $\sqrt{s_{\rm NN}} = 200$ GeV in five centrality classes ($0\%$--$20\%$, $20\%$--$40\%$, $40\%$--$60\%$, $60\%$--$80\%$, and $80\%$--$100\%$). Calculations are performed using the (3+1)D relativistic viscous hydrodynamic program CLVisc with parametric longitudinal entropy deposition initial conditions, freeze-out temperature $T_{\mathrm{frz}} = 128$ MeV, and without afterburner. Experimental data from the PHOBOS collaboration~\cite{PHOBOS:2004fzb} are shown as scattered points. Solid lines represent simulation results, with different colors indicating different centrality classes. Panels (a), (b), (c), and (d) correspond to the four parameter sets in Table~\ref{tab:parameter_for_IC_3D}. Panels (a)--(c) adopt fixed pseudorapidity losses with $\Delta \eta_s^L = 1.36$ and $\Delta \eta_s^R = 4.36$. In contrast, panel (d) employs $\Delta \eta_s^L = 1.36$ and a $\Delta \eta_s^R$ that depends on the number of binary collisions $n_{\mathrm{BC}}$.}
  \label{fig:dNdEta_dis}
\end{figure*}

\begin{figure}[htp]
  \includegraphics[width=0.5 \textwidth]{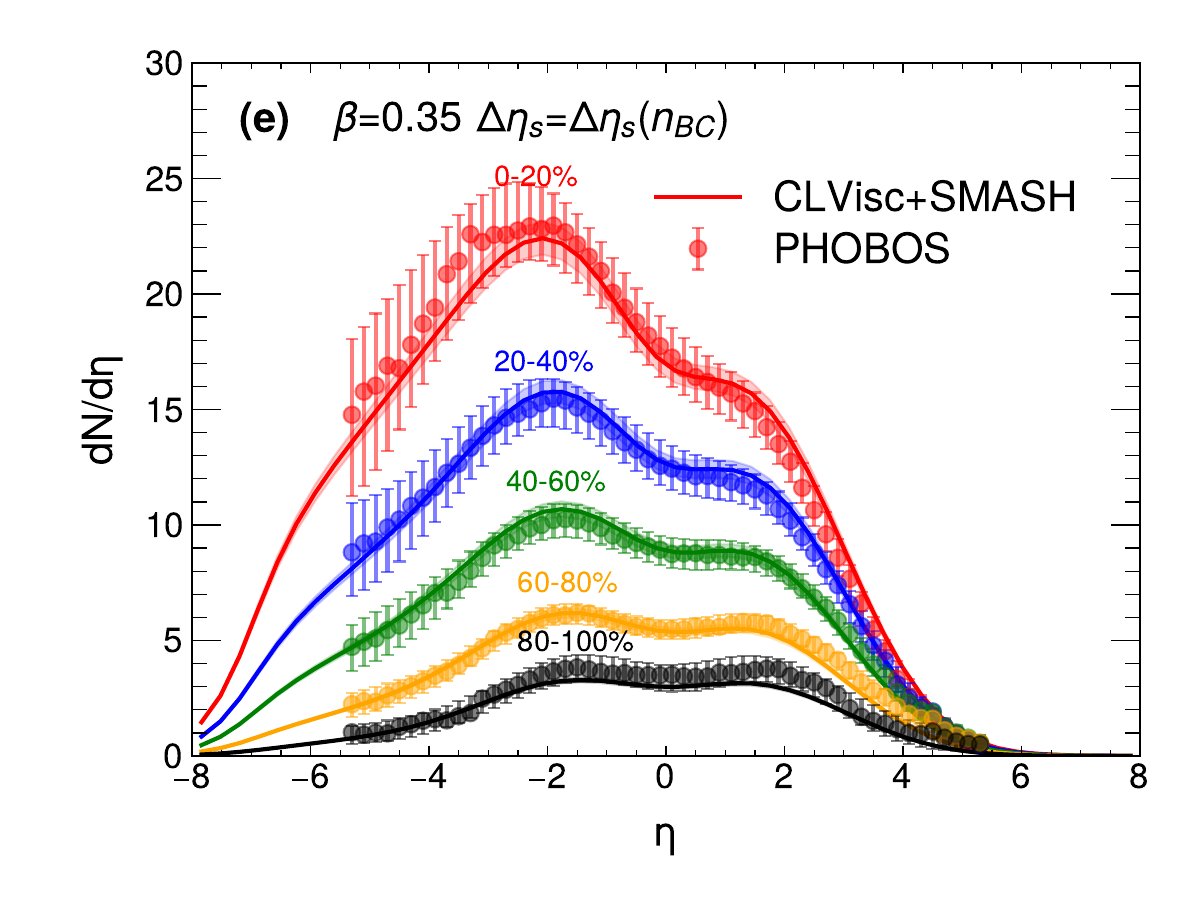}
  \caption{(Color online) Charged particle pseudorapidity distribution $\mathrm{d}N_{\mathrm{ch}}/\mathrm{d}\eta$ for d+Au collisions at $\sqrt{s_{\rm NN}} = 200$ GeV in centrality classes $0\%$--$20\%$, $20\%$--$40\%$, $40\%$--$60\%$, $60\%$--$80\%$, and $80\%$--$100\%$. Results are obtained using the (3+1)D relativistic viscous hydrodynamics program CLVisc and the SMASH model. Different colors indicate different centrality classes; solid lines: simulation results; scattered points: experimental data from the PHOBOS Collaboration~\cite{PHOBOS:2004fzb}. Using parameter set (d) in Table~\ref{tab:parameter_for_IC_3D}, the pseudorapidity loss $\Delta\eta_s^R$ depends on $n_{\mathrm{BC}}$. Hydrodynamic evolution continues to the freeze-out temperature $T_{\mathrm{frz}} = 150$~MeV, followed by hadronization and hadronic rescattering via the SMASH transport mode.}
  \label{fig:dNdEta_dis_after_smash}
\end{figure}

In d+Au collisions, we first perform a rapid scan of various longitudinal entropy deposition scenarios that produce different 3D initial conditions. For this purpose, we evolve the QGP solely with CLVisc at a freeze-out temperature of $T_{\rm frz} = 0.128$ GeV, without employing the SMASH afterburner. This allows us to efficiently identify the optimal parameter set in Table~\ref{tab:parameter_for_IC_3D}.  

Fig.~\ref{fig:dNdEta_dis} compares the simulated charged-particle multiplicity distributions in d+Au collisions at $\sqrt{s_{\rm NN}} = 200$ GeV with PHOBOS data~\cite{PHOBOS:2004fzb} for centrality classes $0\%-20\%$, $20\%-40\%$, $40\%-60\%$, $60\%-80\%$, and $80\%-100\%$. Panels (a)–(d) correspond to the four parameter sets in Table~\ref{tab:parameter_for_IC_3D}.

With parameter set (a) (fixed $\Delta \eta_s^L=1.36$, $\Delta \eta_s^R=4.36$ and $\beta=0.5$), the simulation partially reproduces central collisions ($0\%-20\%$, $20\%-40\%$) but poorly describes peripheral collisions. Set (b) (fixed $\Delta \eta_s^L=1.36$, $\Delta \eta_s^R=4.36$ and $\beta=0.4$) improves both central and peripheral ($40\%-60\%$ and $60\%-80\%$) collisions. Set (c) (fixed $\Delta \eta_s^L=1.36$, $\Delta \eta_s^R=4.36$ and $\beta=0.35$) further enhances agreement across all centralities.

In set (d), we introduce a binary collision number ($n_{\rm BC}$)-dependent $\Delta \eta_s^R$ and $\Delta \eta_s^L=1.36$, while keeping $\beta=0.35$. This further improves the description in the $\eta>2$ region for $0\%-20\%$ collisions. Thus, a collision-number-dependent pseudorapidity loss provides a more accurate physical description, making set (d) the optimal choice for constructing initial conditions in d+Au collisions.

The entropy deposition coefficient $\beta$ and pseudorapidity loss $\Delta \eta$ are key parameters for the final-state distribution. Comparing panels (a)–(c) shows that reducing $\beta$ significantly improves peripheral collisions. Comparing panels (c) and (d) shows that an $n_{\rm BC}$-dependent $\Delta \eta$ further enhances the description in the $\eta>2$ region. These findings provide guidance for optimizing initial conditions and offer new insights into the longitudinal entropy deposition mechanism in d+Au collisions.

\subsection{CLVisc+SMASH Results: Pseudorapidity Distributions, Transverse Momentum Spectra, and Anisotropic Flows}

For a more precise description of the QGP evolution, we adopt the CLVisc+SMASH framework with a freeze-out temperature of $T_{\rm frz} = 0.15$ GeV, utilizing the initial conditions from parameter set (d). 

Fig.~\ref{fig:dNdEta_dis_after_smash} shows the resulting $dN_{ch}/d\eta$ distribution. Compared to graph (d) in Fig.~\ref{fig:dNdEta_dis}, the SMASH afterburner introduces only minor modifications: a slight reduction in the $0$–$20\%$ centrality region at $\eta > 2$ and a marginal enhancement in peripheral collisions, leading to a modest overall improvement. 

Using the same framework and parameter set, we compute the identified particle spectra and anisotropic flows in d+Au collisions and compare them with the PHENIX experimental data~\cite{PHENIX:2013kod,PHENIX:2018lia}. 

\begin{figure*}[ht]
  \includegraphics[width=0.95\textwidth]{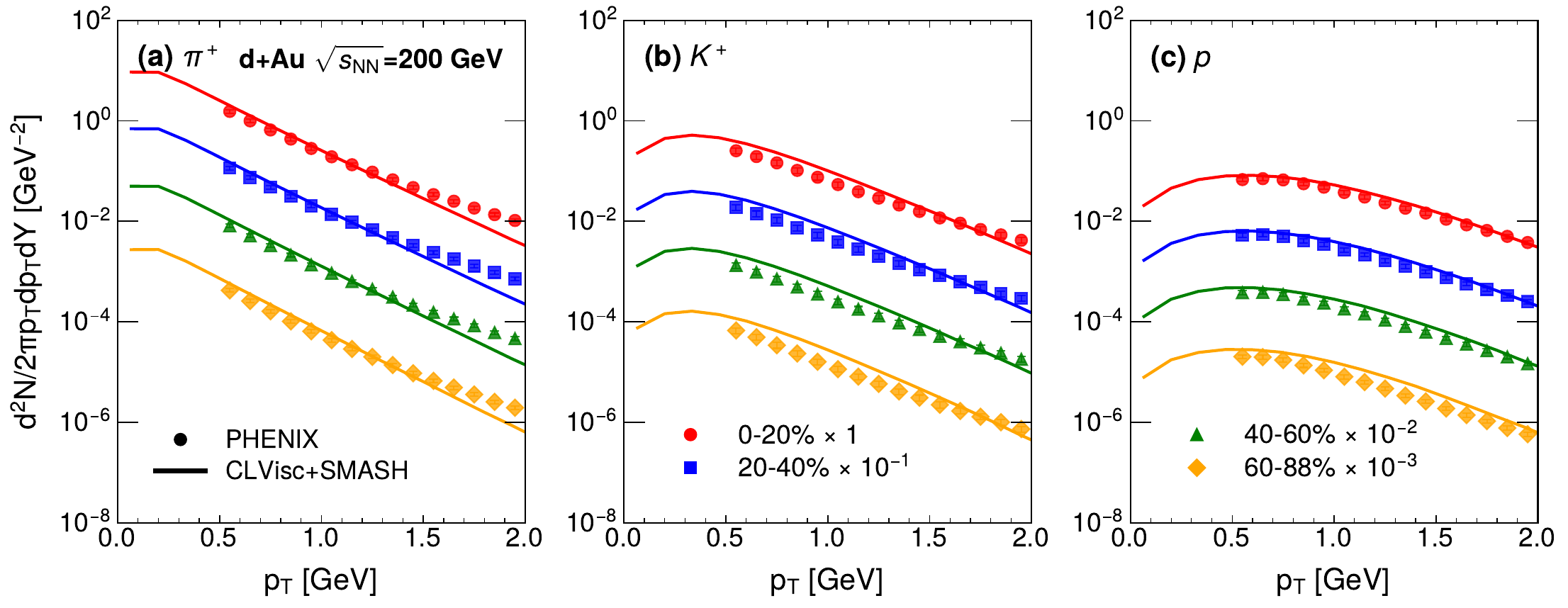}
  \caption{(Color online) Invariant yield of $\pi^+$, $K^+$, and $p$ in d+Au collisions at $\sqrt{s_{\rm NN}} = 200$ GeV for centrality classes $0\%$--$20\%$, $20\%$--$40\%$, $40\%$--$60\%$, and $60\%$--$88\%$. Results are obtained from CLVisc+SMASH simulations using a constant specific shear viscosity $\eta/s = 0.08$, an initial proper time $\tau_0 = 0.6$ fm/$c$, and a freeze-out temperature $T_{\rm frz} = 0.15$ GeV, with initial conditions from parameter set (d). Different colors indicate different centrality classes; solid lines: CLVisc+SMASH results; scattered points: experimental data from the PHENIX Collaboration~\cite{PHENIX:2013kod}.}
  \label{fig:pT_spectra_Tfrz_dAu}
\end{figure*}

Fig.~\ref{fig:pT_spectra_Tfrz_dAu} shows the transverse momentum spectra for $\pi^+$, $K^+$, and $p$. The model results agree reasonably with the data over most of the measured $p_T$ range. However, the $\pi^+$ yield is slightly underestimated at $p_T \gtrsim 1.6$ GeV, a region where the Cooper-Frye direct hadronization alone becomes insufficient. This issue can be addressed by incorporating the quark coalescence mechanism, as demonstrated in Ref.~\cite{Zhao:2020wcd}, where the hybrid Hydro-Coal-Frag model provides a consistent description of the $\pi^+$, $K^+$, and $p$ spectra from low to high $p_T$.

\begin{figure}[htp]
  \includegraphics[width=0.45\textwidth]{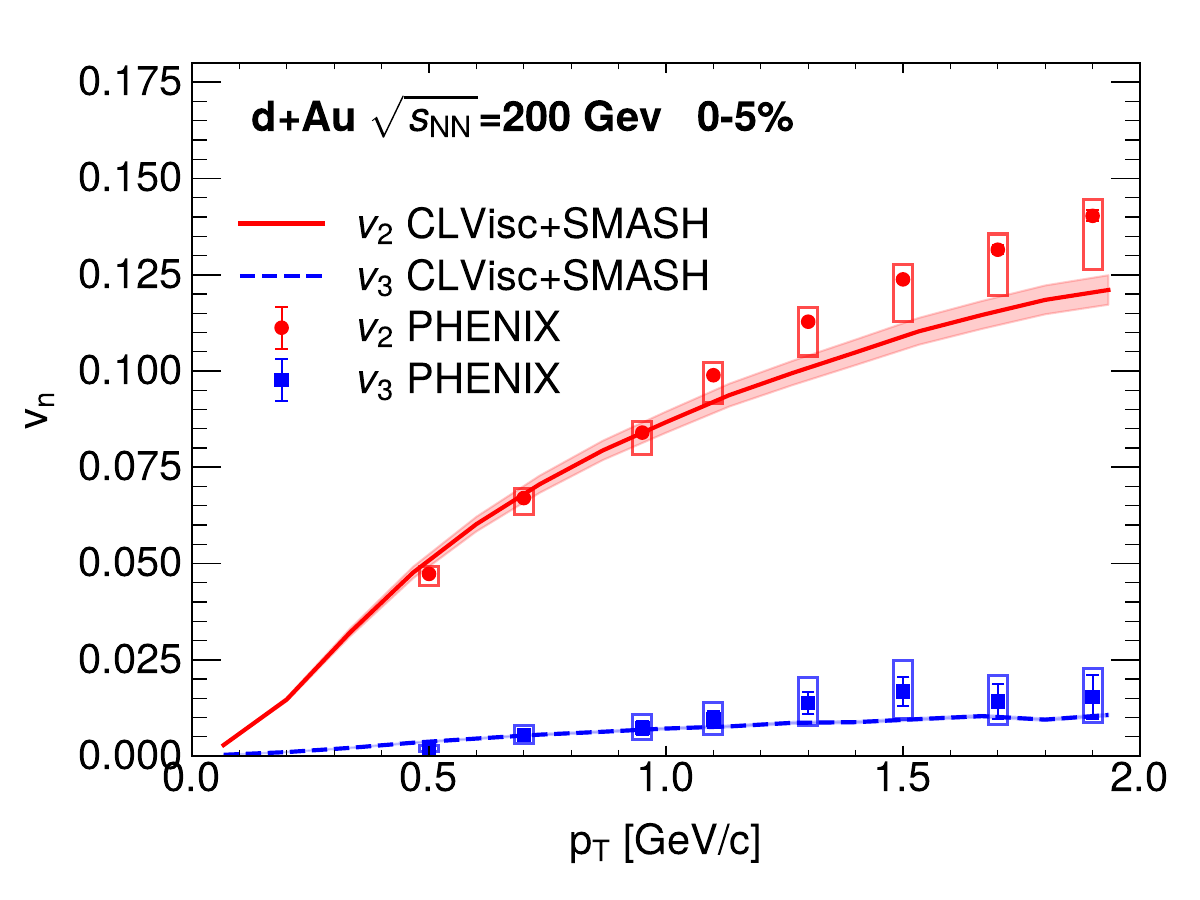}
  \caption{(Color online) The $p_T$ dependence of anisotropic flows $v_2$ and $v_3$ in d+Au collisions at $\sqrt{s_{\rm NN}} = 200$ GeV for centrality $0\%$--$5\%$. Results are obtained using the CLVisc+SMASH framework with a freeze-out temperature $T_{\rm frz} = 0.150$ GeV and $\eta/s = 0.08$, utilizing initial conditions from parameter set (d). Solid lines: simulation results; data points with error bars: experimental measurements from the PHENIX Collaboration~\cite{PHENIX:2018lia}. The present framework, incorporating both hydrodynamic evolution and the SMASH afterburner, effectively describes anisotropic flow observables in small collision systems.}
  \label{fig:vn_pt_dAu}
\end{figure}

Fig.~\ref{fig:vn_pt_dAu} presents the $p_T$ dependence of $v_2$ and $v_3$ for the $0$–$5\%$ centrality class. Our calculations are in good agreement with the experimental data across most of the $p_T$ range, although $v_2$ is slightly underestimated at high $p_T$. This discrepancy may indicate the need to incorporate subnucleon fluctuations in the initial state, which could enhance the high-$p_T$ anisotropic flow~\cite{STAR:2022pfn,Huang:2025cjm}.

\subsection{Test the Universality of the Longitudinal Entropy Deposition Model Using Other Asymmetry Small System}

\begin{figure*}[htp]
  \includegraphics[width=1.0\textwidth]{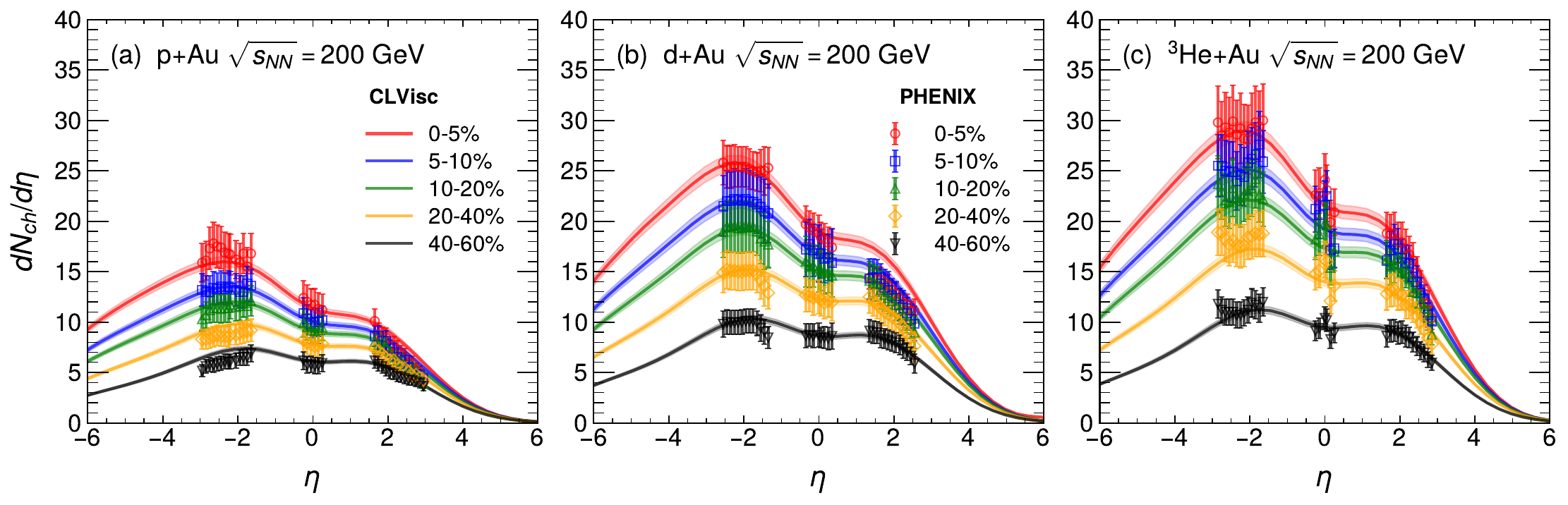}
  \caption{(Color online) Pseudorapidity distributions $\mathrm{d}N_{\mathrm{ch}}/\mathrm{d}\eta$ for p+Au, d+Au and $^3$He+Au collisions at $\sqrt{s_{\rm NN}} = 200$ GeV in five centrality classes ($0\%$--$5\%$, $5\%$--$10\%$, $10\%$--$20\%$, $20\%$--$40\%$, and $40\%$--$60\%$). Calculations are performed using the (3+1)D relativistic viscous hydrodynamic program CLVisc with parametric longitudinal entropy deposition initial conditions, freeze-out temperature $T_{\mathrm{frz}} = 128$ MeV(without afterburner). Experimental data from the PHENIX collaboration~\cite{PHENIX:2018hho} are shown as scattered points}
  \label{fig:dndeta_small_systems}
\end{figure*}

The predictive power of our (3+1)D hydrodynamic model with parametric initial conditions is first assessed in small asymmetric collision systems. The entropy deposition parameters are calibrated using PHOBOS d+Au data at $\sqrt{s_{\rm NN}} = 200$ GeV, yielding the optimized parameter set (d). Without further adjustments, the model is then applied to predict pseudorapidity distributions in p+Au, d+Au and $^{3}$He+Au collisions at the same energy, with results compared to PHENIX measurements. All calculations employ CLVisc with freeze-out temperature $T_{\rm frz} = 128$ MeV, omit the SMASH afterburner for computational efficiency, and follow experimental centrality definitions using the forward rapidity range $-3.9 < \eta < -3.1$\cite{PHENIX:2018hho}. Since our previous results show that the afterburner has a negligible effect on $dN_{ch}/d\eta$ in these small systems, we omit it to enable fast simulations across the three collision geometries. This systematic comparison across multiple small-system geometries provides a stringent test of the initial condition parametrization.

Fig.~\ref{fig:dndeta_small_systems} presents the pseudorapidity distributions $\mathrm{d}N_{\mathrm{ch}}/\mathrm{d}\eta$ for p+Au, d+Au, and $^{3}$He+Au collisions at $\sqrt{s_{\rm NN}} = 200$ GeV across five centrality classes (0\%--5\%, 5\%--10\%, 10\%--20\%, 20\%--40\%, and 40\%--60\%). The CLVisc calculations, employing the parametric longitudinal entropy deposition initial conditions with freeze-out temperature $T_{\rm frz} = 128$ MeV and without afterburner, are compared with PHENIX experimental data\cite{PHENIX:2018hho}. Overall, our hydrodynamic model successfully reproduces the measured pseudorapidity distributions across all three collision systems and centrality bins.

The significance of this result lies in demonstrating the universality of the model. The accurate description of the pseudorapidity distribution, transverse momentum spectra, and flow observables in the Au+Au system (see Figs.~\ref{fig:dNdeta_AuAu}, \ref{fig:pT_spectra_AuAu} and  \ref{fig:Au_Au_vn_pt} in the Appendix), together with its success in asymmetric collision systems, confirms the robustness and generality of our parametrization. This provides strong support for a unified hydrodynamic description across different collision systems, provided that a correct longitudinal entropy deposition mechanism is employed.

\section{Summary and discussion}\label{sec:summary}

Previous studies have shown that describing the final-state $dN_{\rm ch}/d\eta$ distribution in asymmetric  collisions across all centrality classes remains challenging when using existing initial condition models with hydrodynamic simulations. To address this, we propose two potential solutions: one considering nuclear structure effects, and the other emphasizing the initial-state longitudinal entropy deposition mechanism.

For the nuclear structure, we examined two forms of the deuteron: the DWF and HWF forms. Our analysis shows that geometric eccentricity-related quantities in the DWF form exhibit lower values as a function of multiplicity compared to the HWF form. This discrepancy likely arises from the more compact nucleon distribution in the DWF form, suggesting that nuclear structure information is retained in the initial-state geometric observables of heavy-ion collisions. However, this information is not reflected in the calculation of the initial longitudinal entropy density distribution, indicating that considering the deuteron structure alone is insufficient to resolve the $dN_{\rm ch}/d\eta$ distribution issue.

For the longitudinal entropy deposition, we proposed a new parameterized formula taking into account the entropy deposition parameter $\beta$ and $n_{\rm BC}$ dependent pseudorapidity loss. Our formula incorporates two key considerations: (1) the possibility that not all energy from the central fireball is converted into final-state particles in small-system collisions, which we explored by varying the entropy deposition parameter $\beta$; and (2) the inclusion of a pseudorapidity loss term, $\Delta \eta_s$, which depends on the number of binary collisions $n_{\rm BC}$. These modifications significantly improved the fit of the $\rm dN_{ch}/d\eta$ distribution. 

Our findings indicate that while the deuteron structure does not significantly aid in resolving the $dN_{\rm ch}/d\eta$ distribution, nuclear structure information can be retained in the geometric observables of the initial state. In small-system collisions, a conversion factor of $\beta \approx 1/3$ may be more appropriate, whereas in Au+Au collisions, $\beta = 1/2$, suggesting that smaller systems require a smaller entropy deposition coefficient. Additionally, pseudorapidity loss plays a critical role in the collision dynamics.

We further computed the transverse momentum spectra of identified particles and anisotropic flow coefficients, achieving good agreement with experimental data. The entropy deposition model employed in this work demonstrates strong universality: it simultaneously provides a satisfactory description of the $dN_{ch}/d\eta$ distribution, identified particle spectra, and anisotropic flow $v_n$ in Au+Au collisions at $\sqrt{s_{\rm NN}}=200$ GeV as well as  $dN_{ch}/d\eta$ in p+Au and $^3$He+Au collisions.

It is important to acknowledge the limitations of our work. Our longitudinal entropy deposition framework is phenomenological, with parameters determined by fitting to the experimental data of d+Au collisions. Nevertheless, it demonstrates good universality when applied to other collision systems at $\sqrt{s_{\rm NN}}=200$ GeV.

In d+Au collision systems, comprehensive experimental investigations have been performed to study multiple physical phenomena. 
These include midrapidity transverse energy distributions $dE_T/d\eta$~\cite{PHENIX:2013ehw}, nuclear modification factors $R_{dAu}$ measurements~\cite{BRAHMS:2004xry,PHENIX:2004nzn,Citron:2009eu,PHENIX:2013jxf}, and detailed analyses of nuclear effects in high-$p_T$ momentum spectra~\cite{Vitev:2002pf,BRAHMS:2003sns}. Furthermore, significant research efforts have focused on understanding the Cronin effect and energy loss mechanisms in nucleon-gold interactions~\cite{Reed:2006aut}, as well as azimuthal correlation patterns~\cite{PHENIX:2013ktj,Sickles:2014lra}. 
Jet production characteristics and initial-state nuclear effects on jet formation have also been systematically examined~\cite{PHENIX:2005veb,Kapitan:2010ep,Kapitan:2009dz}. 
Additionally, studies have explored forward $\Lambda$ production, low-$p_T$ $J/\Psi$ production mechanisms and collective flow and the fluid behaviors, providing valuable insights into the complex dynamics of d+Au collision systems~\cite{STAR:2007lcq,STAR:2016uxt,Wu:2023vqj,Shen:2016zpp}. 
In the future, our well-calibrated d+Au simulations can be extended to these studies.

The current analysis focuses on basic observables, including $dN_{\rm ch}/d\eta$, identified particle spectra, and anisotropic flow. Future work could extend to calculating longitudinal correlations and decorrelation effects. Additionally, the applicability of our proposed formula could be tested in other small-system collisions, such as $p+\mathrm{Pb}$, $\mathrm{O}+\mathrm{O}$, Ne+Ne, and Ne+Pb collisions, providing further constraints on the longitudinal entropy deposition as well as nuclear structure information.


\begin{acknowledgments}

This work is supported by the National Natural Science Foundation of China under Grant No.\ 12075098, No.\ 12435009, and No.\ 12575140. The work is also partly supported by the Outstanding Leading Talent Team Program of Central China Normal University (XJ2026000302). The numerical calculation have been performed on the GPU cluster in the Nuclear Science Computing Center (NSC3) at CCNU. 

\end{acknowledgments}

\begin{appendix}
\section{Test the Universality of the Longitudinal Entropy Deposition Model Using Symmetric Au+Au Collision}

We extend the same framework to symmetric Au+Au collisions at $\sqrt{s_{\rm NN}} = 200$ GeV. The baseline parameter set (d) is retained, while the entropy deposition coefficient is optimized to $\beta = 0.5$ to accommodate the larger system size and higher energy density. As in the small-system cases, the evolution is simulated solely with CLVisc ($T_{\rm frz} = 128$ MeV) without afterburner. Centrality definitions follow experimental conventions: $3 < |\eta| < 4.5$ for $\mathrm{d}N_{ch}/\mathrm{d}\eta$~\cite{Back:2002wb}, $3.0 < |\eta| < 3.9$ for identified particle spectra~\cite{PHENIX:2003iij}, and $3.1 < |\eta| < 3.9$ for anisotropic flow~\cite{PHENIX:2011yyh}. 

For Au+Au collisions, both left- and right-going nuclei experience multiple scatterings. We therefore extend the $n_{\rm BC}$-dependent rapidity loss parametrization to both sides, in contrast to the d+Au case where only the deuteron side carries this dependence. Specifically, we adopt the same functional form as defined in Eq.~(\ref{eq:delta_eta}) for both $\Delta\eta_s^L(n_{\rm BC})$ and $\Delta\eta_s^R(n_{\rm BC})$, with the constant baseline values $\Delta\eta_s^L = \Delta\eta_s^R = 4.36$.

\begin{figure}[htp]
    \centering
    \includegraphics[width=0.45\textwidth]{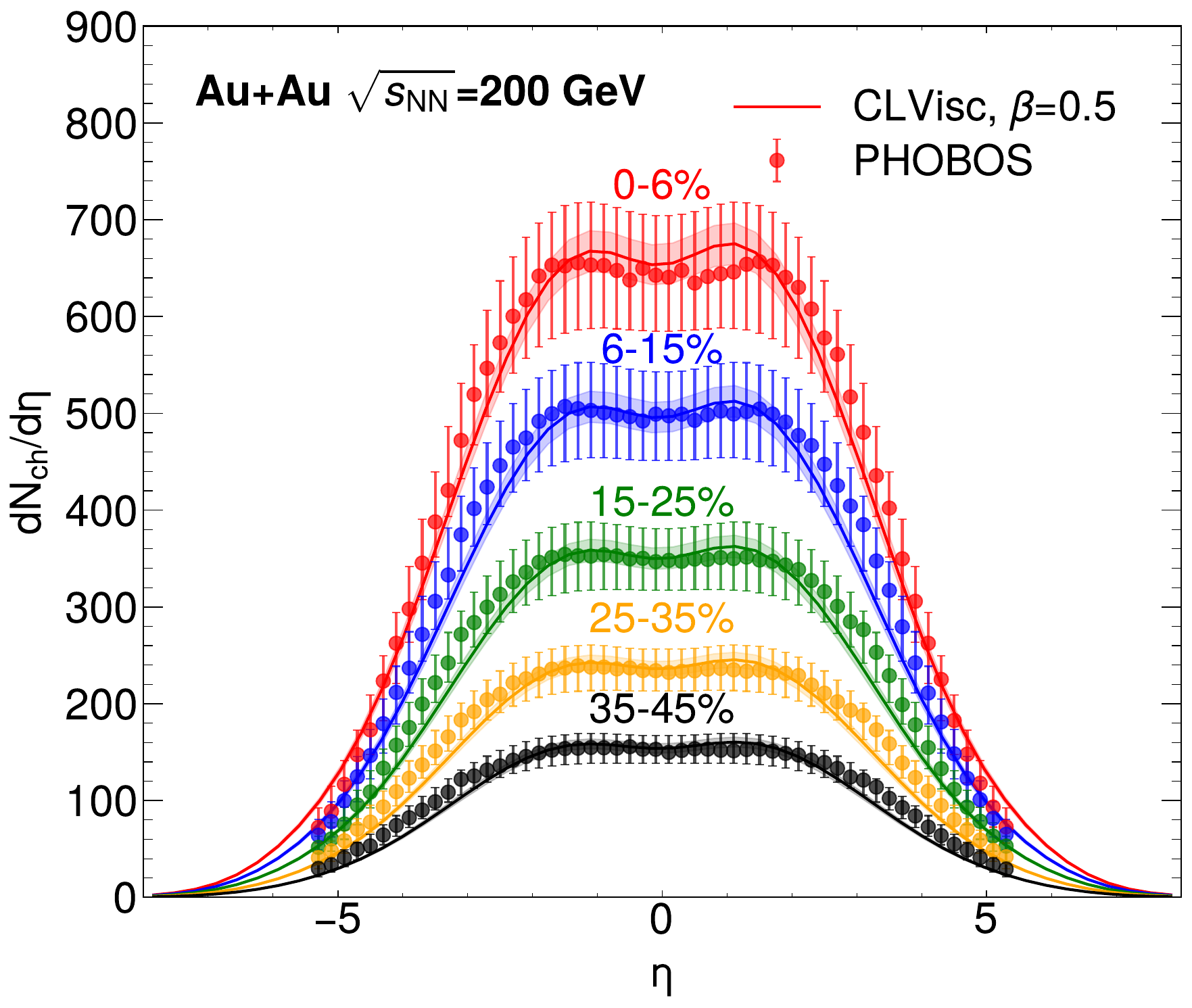}
    \caption{(Color online) Charged particle multiplicity distributions $\mathrm{d}N_{\mathrm{ch}}/\mathrm{d}\eta$ in Au+Au collisions at $\sqrt{s_{\rm NN}} = 200$ GeV for centrality classes $0\%$--$6\%$, $6\%$--$15\%$, $15\%$--$25\%$, $25\%$--$35\%$, and $35\%$--$45\%$. The initial conditions are generated using parameter set (d) in Table~\ref{tab:parameter_for_IC_3D} with $\beta = 0.5$, and the system is evolved with the (3+1)D relativistic viscous hydrodynamic code CLVisc using a freeze-out temperature of $128$ MeV. Different colors indicate different centrality classes. Solid lines represent simulation results, and the shaded bands indicate the corresponding statistical uncertainties. Data points with error bars are experimental measurements from the PHOBOS Collaboration~\cite{Back:2002wb}.}
    \label{fig:dNdeta_AuAu}
\end{figure}

\begin{figure*}[htp]
    \includegraphics[width=0.9\textwidth]{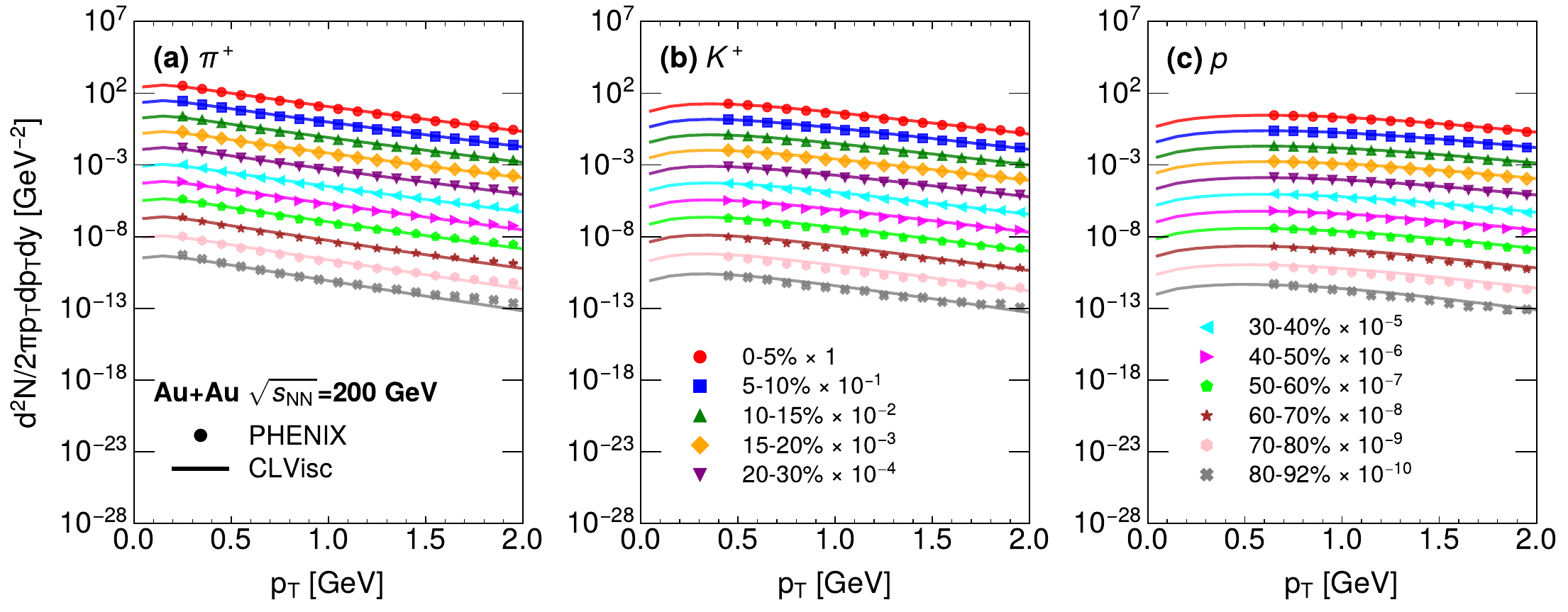}
    \caption{(Color online) Transverse momentum spectra of identified particles ($\pi^+$, $K^+$, $p$) in Au+Au collisions at $\sqrt{s_{\rm NN}} = 200$ GeV for centrality classes $0\%$--$5\%$, $5\%$--$10\%$, $10\%$--$15\%$, $15\%$--$20\%$, $20\%$--$30\%$, $30\%$--$40\%$, $40\%$--$50\%$, $50\%$--$60\%$, $60\%$--$70\%$, $70\%$--$80\%$, and $80\%$--$92\%$. Results are obtained from (3+1)D viscous hydrodynamic simulations using CLVisc with a freeze-out temperature $T_{\rm frz} = 128$ MeV, shear viscosity-to-entropy density ratio $\eta/s = 0.08$, and initial proper time $\tau_0 = 0.6$ fm/$c$. Different colors indicate different centrality classes; solid lines: CLVisc simulation results; scattered symbols with error bars: experimental data from the PHENIX Collaboration~\cite{PHENIX:2003iij}. }
    \label{fig:pT_spectra_AuAu}
\end{figure*}

\begin{figure*}[htp]
    \centering
    \includegraphics[width=0.95\textwidth]{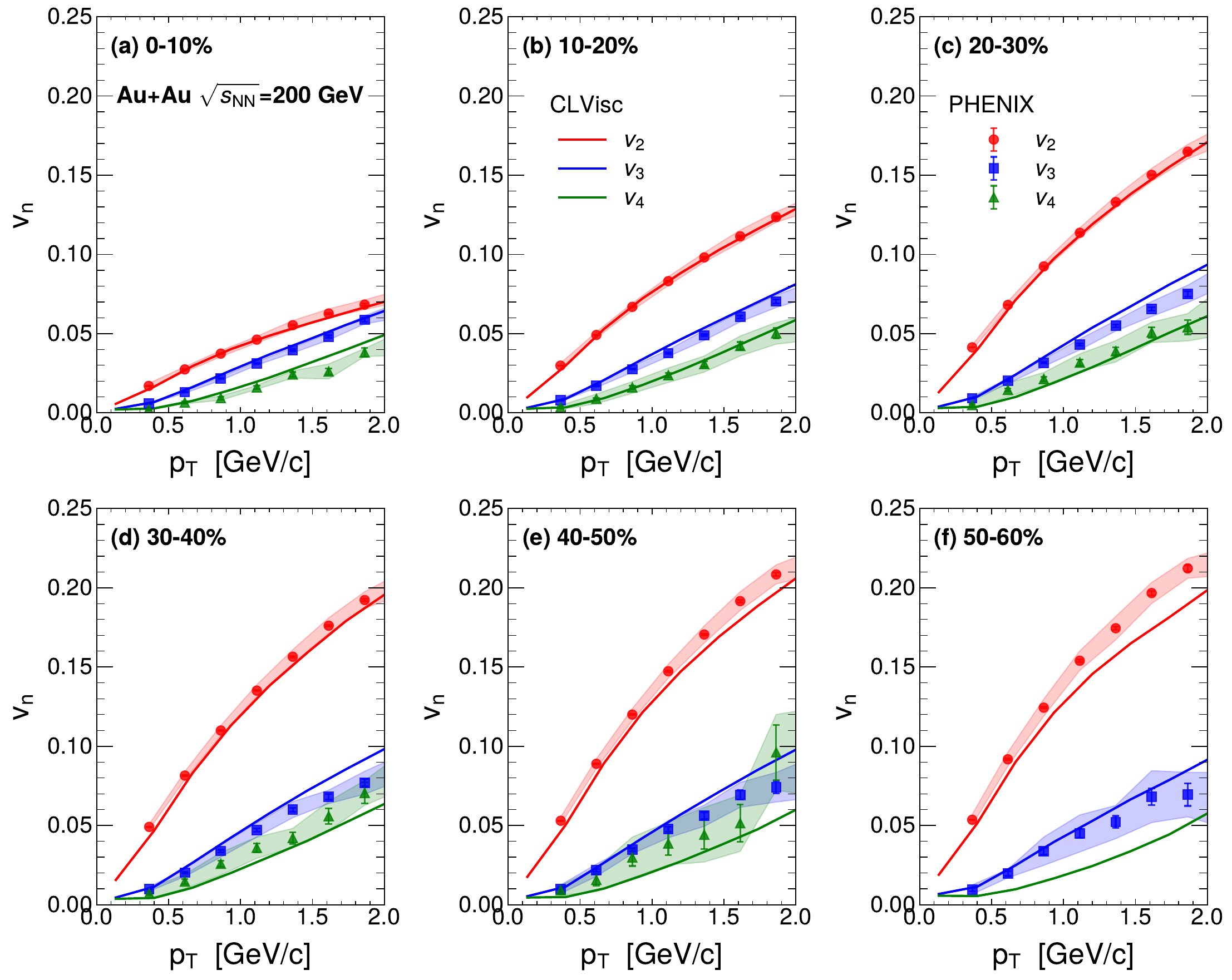}
    \caption{(Color online) Centrality dependence of anisotropic flows $v_n$ ($n=2,3,4$) in Au+Au collisions at $\sqrt{s_{\rm NN}} = 200$ GeV for centrality bins $0\%$--$10\%$, $10\%$--$20\%$, $20\%$--$30\%$, $30\%$--$40\%$, $40\%$--$50\%$, and $50\%$--$60\%$. Results are obtained from simulations using the (3+1)D relativistic viscous hydrodynamic code CLVisc with $T_{\rm frz} = 128$ MeV and $\eta/s = 0.08$. Different colors indicate different $v_n$ ($n=2,3,4$); solid lines: CLVisc simulation results; data points with error bands: experimental measurements from the PHENIX Collaboration~\cite{PHENIX:2011yyh}. The results demonstrate that the longitudinal entropy distribution effectively describes the generation and evolution of anisotropic flows even in symmetric systems. }  
    \label{fig:Au_Au_vn_pt}
\end{figure*}

Fig.~\ref{fig:dNdeta_AuAu} shows the pseudorapidity distributions for Au+Au collisions at $\sqrt{s_{\rm NN}}=200$ GeV for five centrality classes: 0--6\%, 6--15\%, 15--25\%, 25--35\%, and 35--45\%. The centrality dependence of the event-averaged charged multiplicity is determined by the event-by-event distributions of the initial total entropy. The CLVisc simulations are compared with PHOBOS data, showing satisfactory agreement across all centrality classes. 

Fig.~\ref{fig:pT_spectra_AuAu} compares the transverse momentum spectra for $\pi^+$, $K^+$, and $p$ from CLVisc simulations with PHENIX data across central to peripheral Au+Au collisions. The model provides a good description of the data over the entire centrality and $p_T$ ranges.

Fig.~\ref{fig:Au_Au_vn_pt} depicts the transverse momentum dependence of anisotropic flows $v_2$, $v_3$, and $v_4$ in Au+Au collisions at $\sqrt{s_{\rm NN}}=200$ GeV across centrality ranges from 0\%–10\% to 50\%–60\%. Our CLVisc simulations show excellent agreement with the experimental data over the entire $p_T$ and centrality ranges, validating the model's description of collective expansion.
\end{appendix}

\bibliography{ref}

\end{document}